\documentclass[aps,twocolumn,amssymb,amsmath,showpacs,a4paper,superscriptaddress]{revtex4-2}
\usepackage{graphicx,color}
\usepackage{amsfonts}
\usepackage{mathtools}
\usepackage{bm}
\newcommand{\grad}{\bm{\nabla}}
\newcommand{\m}{\widetilde{m}}
\newcommand{\3}{\mathrm{3D}}
\newcommand{\2}{\mathrm{2D}}

\begin{document}

\title{Stability of quantised vortices in two-component condensates}

\author{Sam Patrick}
\email{samuel\_christian.patrick@kcl.ac.uk}
\affiliation{Department of Physics, King’s College London, The Strand, London WC2R 2LS, United Kingdom}

\author{Ansh Gupta}
\email{ansh.gupta@kcl.ac.uk}
\affiliation{Department of Physics, King’s College London, The Strand, London WC2R 2LS, United Kingdom}

\author{Ruth Gregory}
\email{ruth.gregory@kcl.ac.uk}
\affiliation{Department of Physics, King’s College London, The Strand, London WC2R 2LS, United Kingdom}
\affiliation{Perimeter Institute, 31 Caroline St, Waterloo, Ontario N2L 2Y5, Canada}

\author{Carlo F. Barenghi}
\email{carlo.barenghi@newcastle.ac.uk}
\affiliation{Joint Quantum Centre Durham-Newcastle, School of Mathematics, Statistics and Physics, Newcastle University, Newcastle upon Tyne NE1 7RU, United Kingdom}

\date{\today}

\begin{abstract}
\noindent
Multiply quantised vortices (MQVs) within single component Bose-Einstein condensates are unstable and decay rapidly. 
We show that MQVs can be stabilised by adding a small number of atoms of a second species to the vortex cores, and that these atoms remain in the vortex core as the system evolves. 
A consequence of the stabilisation is that nearby co-rotating vortices can orbit in the opposite sense to their individual rotations when enough of the second species is present.
This has implications concerning the imaging of vortices, as well as quantum turbulence and vortex nucleation in two-component condensates, such as those involving mixtures of $^{87}$Rb and $^{133}$Cs.
\end{abstract}

\maketitle

\section{Introduction}

The most striking property of quantum fluids (superfluid ${}^4$He and ${}^3$He, atomic Bose-Einstein condensates, polariton condensates, etc) is the existence of a macroscopic complex wavefunction $\Psi(\mathbf{x},t)$, where $\mathbf{x}$ is the position and $t$ is time. If we write the wavefunction as $\Psi = |\Psi|e^{i\Phi}$ in terms of its amplitude $|\Psi|$ and phase $\Phi$, then the density $n$ and the velocity $\mathbf{v}$ of the superfluid are \cite{barenghi2016primer}:
\begin{equation}
    n(\mathbf{x},t) = |\Psi(\mathbf{x},t)|^2, \qquad \mathbf{v}(\mathbf{x},t) = \frac{\hbar}{M}\grad\Phi(\mathbf{x},t),
\end{equation}
where $\hbar$ is the reduced Planck’s constant and $M$ is the mass of the relevant boson. The wavefunction imposes constraints on the rotational motion.
Whereas rotation in ordinary (classical) fluids takes the form of eddies of arbitrary size and circulation, rotation in superfluids is quantised, occurring around special points (in two-dimensions) or lines (in three-dimensions) called quantised vortices. 
At these special points the density $n$ is exactly zero, hence the phase $\Phi$ is not defined: quantised vortices are therefore phase defects. 
Within a superfluid, the circulation $\Gamma$ of the velocity along a closed path $C$ is either zero if the region inside $C$ does not include any vortex, or an integer multiple $\ell$ of the quantum of circulation $\kappa= 2\pi \hbar/M$ \cite{barenghi2016primer}, 
\begin{equation}\label{kappa}
    \Gamma = \oint \mathbf{v}\cdot d\mathbf{r} = \ell\kappa,
\end{equation}
if the region inside $C$ includes a vortex of quantisation $\ell$ (since $\Psi$ must be continuous, $\ell$ must take on integer values). This condition strongly constrains the velocity field around the vortex axis.
Vortices with $|\ell|=1$ are commonly referred to as singly quantised vortices, SQVs, whereas $|\ell|>1$ are called multiply quantised vortices, MQVs ($\ell$ is commonly called the ``charge'' of the vortex).
Since the angular momentum and the kinetic energy carried by a vortex are proportional to $\ell$ and $\ell^2$
respectively, MQVs are energetically unstable \cite{adhikari2019stable,adhikari2002effect,srinivasan2006vortices} in simple homogeneous superfluids \cite{kuopanportti2015ground};
any MQV with charge $\ell$ will rapidly decay into $\ell$ SQVs carrying the same total angular momentum.
This decay is the result of an energetic instability in the spectrum of the vortex, i.e. the vortex possesses a normal mode with negative energy.
In the presence of a dissipation mechanism, the amplitude of this mode will grow eventually causing the central vortex to split up.
In large enough systems, the surrounding bath of sound waves can also act as a reservoir for the vortex to dissipate energy into, allowing MQVs to split even when the governing equations are energy conserving.
In this case, the spectrum of the vortex contains an unstable mode with complex frequency which grows spontaneously, see e.g. \cite{pu1999coherent,shin2004dynamical,isoshima2007spontaneous,giacomelli2020ergoregion,patrick2022origin,patrick2022quantum}.

The possibility of stabilizing inherently unstable MQVs has been explored by using inhomogeneous trapping potentials to force the spatial distribution of superfluid matter \cite{adhikari2019stable} and by manipulating the superfluid via electro–magnetic field fluctuations \cite{qin2016stable}. 
A third possibility is based on two–component condensates \cite{myatt1997production,hall1998dynamics}. These systems can consist of different atoms \cite{mccarron2011dual}, different isotopes of the same atom \cite{papp2008tunable} or even different hyperfine states of the same isotope \cite{kronjager2005evolution,tojo2010controlling}. 
It has been suggested that in such systems \cite{kuopanportti2019splitting,mason2017solid,zhang2016exotic,dong2016equilibrium,ma2016two,kawaguchi2004splitting} the overlapping second component can slow down or even prevent the decay of MQVs \cite{kuopanportti2015ground} of the first component. As we shall see, spatial separation is relevant to our problem. Two distinct modes of spatial separation exist \cite{ao1998binary,ho1996binary}: separation arising from the geometry of the trapping potentials, and separation resulting from the immiscibility of the two components \cite{timmermans1998phase}, a property characterised by the inter–component interactions. Kuopanportti et al. \cite{kuopanportti2015ground} suggested how to stabilise MQVs in a tightly confined two–dimensional ${}^{87}$Rb and ${}^{41}$K mixture confined within a rotating harmonic trap. 
They successfully stabilised $\ell=2$ MQVs with both attractive and repulsive inter–component interactions by a clever distribution of the secondary component. In each case, the secondary component was distributed in such a way that the energetic favourability of MQV decay was not sufficient to overcome the inter–component interactions: the MQVs were effectively ‘propped up’ and could
not decay. Unfortunately, the scope of this technique is limited by the tightness of the trapping potential: the vortex core was comparable in size to the size of the entire condensate.

In this work, we propose a method of stabilising MQVs which exploits the spatial separation of mixtures with highly repulsive inter–component interactions. We shall show that, by letting the second component concentrate inside the vortex core of the first component, the energetic advantage of vortex decay can be offset. To best facilitate stabilisation of MQVs, it is paramount to choose a condensate mixture such that both components are self-repulsive and the inter-component repulsion is large. A mixture of $^{87}$Rb and $^{133}$Cs atoms meets both requirements and is realisable in the laboratory \cite{mccarron2011dual}. Rather than a conventional harmonic trap, we consider the problem in a potential trap with a wide, flat bottom and steep walls, which allows for condensates that are homogeneous in the bulk \cite{wang2010spin,wu2011unconventional}. The advantage of such large circular bucket potentials is that the vortex core profiles remain unchanged as they move about the condensate, unlike what happens in a tightly bound system, making our results independent of the trap’s details.

\section{The system}

\subsection{Governing equations}

The equations of motion of a trapped two-dimensional two-component condensate are,
\begin{equation} \label{GPE}
\begin{split}
i\hbar\partial_t\Psi_1 = & \ \left[{\hat K}_1 + V(\mathbf{x}) + G_1|\Psi_1|^2 + G_{12}|\Psi_2|^2 - {\tilde \mu_1} \right]\Psi_1, \\
i\hbar\partial_t\Psi_2 = & \ \left[{\hat K}_2 + V(\mathbf{x}) + G_2|\Psi_2|^2 + G_{12}|\Psi_1|^2-{\tilde \mu_2} \right]\Psi_2,
\end{split}
\end{equation}
where $\Psi_j$ are the macroscopic wavefunctions for the components $j=1,2$, ${\tilde \mu_j}$ are the chemical potentials, ${\hat K}_j=-\frac{\hbar^2}{2M_j}\nabla^2$ are the kinetic energy operators, $G_j$ are the 2D self-interaction parameters and $G_{12}$ is the 2D inter-species interaction parameter.
The 2D trapping potential $V(\mathbf{x})$ (where $\mathbf{x}=(x,y)$) is taken to be a cylindrical box-trap of radius $r_B$ and height $V_0$. 
It is convenient to assume the circularly symmetric form,
\begin{equation}
    V(\mathbf{x})=V(r)=\frac{V_0}{1+(V_0-1)e^{a(r_B-r)}},
\end{equation}
where $r^2=x^2+y^2$. Assuming that $V_0$ is larger than both chemical potentials, the two species are essentially confined in the region $r<r_B$; within this region, $V(r)$ is negligible. The quantity $a$ is a smoothing parameter.
We assume the system is sufficiently low temperature that thermal noise and dissipation can be neglected.
The assumption of an effectively two-dimensional system can be satisfied in practice by imposing a strong confinement of both components in the vertical direction, e.g. a harmonic potential with frequency $\omega_z\gg\tilde{\mu}_j/\hbar$, such that the dynamics in the extra dimension are frozen out (see Appendix~\ref{app:dim}).

The system of equations (\ref{GPE}) can be derived from the following action,
\begin{equation} \label{action}
    \mathcal{S} = \int dt d^2\mathbf{x}\left[ \sum_j \left(i\hbar\Psi_j^*\dot{\Psi}_j - \mathcal{H}_j\right) - \mathcal{H}_I\right],
\end{equation}
where the free energy density for each component $j$ and inter-species interaction energy density are respectively defined as,
\begin{equation}
\begin{split}
    \mathcal{H}_j = & \ \Psi_j^*\left[\hat{K}_j+V(\mathbf{x})-{\tilde \mu_j}\right]\Psi_j + \frac{G_j}{2}|\Psi_j|^4, \\
    \mathcal{H}_I = & \ G_{12}|\Psi_1|^2|\Psi_2|^2.
\end{split}
\end{equation}
Since independent phase rotations of $\Psi_1$ and $\Psi_2$ are symmetries of $\mathcal{S}$, we obtain two conserved currents
\begin{equation}
\partial_t n_j + \grad\cdot(n_j\mathbf{v}_j) = 0,
\end{equation}
where $n_j=|\Psi_j|^2$ is the number density of component $j$ and $\mathbf{v}_j=\frac{\hbar}{M_j}\grad\mathrm{arg}(\Psi_j)$ is its velocity.
In a closed system, one can integrate over space to find that the number of atoms in each species, $N_j = \int n_jd^2\mathbf{x}$, is conserved by the evolution dictated by \eqref{GPE}.
Hereafter, we fix the number of atoms in the first component $N_1$ and treat $\eta=N_2/N_1$ as our variable.

An approximate expression for the ground state solution of equations (\ref{GPE}) is provided by the Thomas-Fermi approximation: neglecting the kinetic energy terms, within the trap (where $V(r) \simeq 0$) the steady state solutions of (\ref{GPE}) satisfy,
\begin{equation}
    G_1n_1 + G_{12}n_2 - {\tilde \mu_1} = 0, \quad G_2n_2 + G_{12}n_1 - {\tilde \mu_2} = 0.
\end{equation}
Since the two components are immiscible for $G_{12}>\sqrt{G_1 G_2}$ \cite{timmermans1998phase}, we can assume that component 2 is confined to the  region $A_2=\pi r_0^2$ and component 1 occupies the remaining space $A_1=\pi(r_B^2-r_0^2)$.
We then have $n_1=N_1/A_1$ and $n_2=N_2/A_2$ so that the chemical potentials are approximately,
\begin{equation} \label{TF}
    \tilde{\mu}_1 = \frac{G_1N_1}{A_1}, \qquad \tilde{\mu}_2 = \frac{\eta G_2N_1}{A_2}.
\end{equation}
In this work, we consider systems with a vortex in $\Psi_1$ and no vortex in $\Psi_2$.
In the stationary state, solutions of \eqref{GPE} are of the form,
\begin{equation} \label{stat}
\Psi_1(r,\theta) = \sqrt{n_1(r)}e^{i\ell\theta}, \qquad \Psi_2(r) = \sqrt{n_2(r)},
\end{equation}
where $\ell$ is the winding number of the vortex.
%The winding number is constrained to be an %integer as a result of periodicity %requirements on $\Psi_1$ in the $\theta$ %direction.
%This is referred to as the quantisation or %the charge of the vortex. [already said]
The corresponding velocity field is $\mathbf{v}_1 = \frac{\hbar\ell}{M_1 r}\hat{\mathbf{e}}_\theta$ where $\hat{\mathbf{e}}_{\theta}$ is the unit vector in the azimuthal direction.
We focus on the case $\ell=2$, however, we expect our conclusions to extend to vortices of higher winding numbers.

\subsection{Dimensionless variables}

In the rest of the paper we express the governing equations \eqref{GPE} in dimensionless form using the following characteristic units of length, time and density,
\begin{equation}
    \xi_1 = \frac{\hbar}{\sqrt{{\tilde \mu_{1,0}} M_1}}, \qquad \tau_1 = \frac{\hbar}{{\tilde \mu_{1,0}}}, \qquad n_0=\frac{{\tilde \mu_{1,0}}}{G_1},
\end{equation}
where $\tilde{\mu}_{1,0}=\tilde{\mu}_1(\eta=0)$ is the chemical potential of species 1 in the vortex state when species 2 is absent
In other words, we perform the following rescalings,
\begin{equation} \label{rescale1}
    \frac{r}{\xi_1}\to r, \quad \frac{t}{\tau_1} \to t, \quad \frac{V}{\tilde{\mu}_{1,0}}\to V, \quad \frac{\Psi_j}{n_0^{1/2}}\to\Psi_j,
\end{equation}
so that \eqref{GPE} becomes,
\begin{equation} \label{GPE2}
\begin{split}
i\partial_t\Psi_1 = & \ \left[-\frac{1}{2}\nabla^2 + V(\mathbf{x}) + |\Psi_1|^2 + g_{12}|\Psi_2|^2 - \mu_1\right]\Psi_1, \\
i\partial_t\Psi_2 = & \ \left[-\frac{1}{2m_2}\nabla^2 + V(\mathbf{x}) + g_2|\Psi_2|^2 + g_{12}|\Psi_1|^2-\mu_2\right]\Psi_2,
\end{split}
\end{equation}
where we have defined the new dimensionless parameters,
\begin{equation}
    \mu_j=\frac{\tilde{\mu}_j}{\tilde{\mu}_{1,0}}, \quad g_2=\frac{G_2}{G_1}, \quad g_{12}=\frac{G_{12}}{G_1}, \quad m_2 = \frac{M_2}{M_1}.
\end{equation}
Note the true total number of atoms can be recovered from $N_j=\int d^3\mathbf{x}|\Psi_j|^2$ in these units by multiplying by a factor of $\hbar^2/M_1G_1=d_1/\sqrt{8\pi}a_1$, where $d_1$ is the width of species 1 in the vertical direction and $a_1$ is its scattering length (see Appendix~\ref{app:dim}).

Since $\xi_1\ell$ (where $\xi_1$ is the healing length of the condensate 1 for $\eta=0$) is the characteristic core size of an $\ell$-vortex, in the new units the vortex core now has radius $\sim\ell$.
Furthermore, since the density of condensate 1 approaches $n_0$ far from the vortex axis where $n_2\to 0$, in the new units the bulk value of $n_1$ will approach $\mu_1$.

\subsection{Parameters} \label{sec:param}

In these units, the problem depends only on $m_2$, $g_2$, $g_{12}$, the parameters in the trapping potential and the chemical potentials $\mu_j$.
The latter are related to the total number of atoms in the ground state (see next section).
The s–wave scattering lengths for the ${}^{87}$Rb and ${}^{133}$Cs
mixture \cite{pattinson2013equilibrium} are $a_1 = 100a_0$, $a_2 = 280a_0$ and $a_{12} = 650a_0$
[26] where $a_0$ is Bohr radius and hereafter the subscripts $1$ and $2$ refer respectively to the ${}^{87}$Rb and ${}^{133}$Cs components. 
In Appendix~\ref{app:dim}, we show that when both species are subject to the same strong vertical confinement, the dimensionless interaction parameters are $g_2=2.04$ and $g_{12}=5.65$.
The mass ratio is simply $m_2=133/87$.
For the trapping potential, we fix $a=5$ and $V_0=5$ so that there is essentially a hard wall located at $r=r_B$.
For the trap radius, we pick $r_B=40$ so that the trap is not too small to prohibit the instabilities under consideration but not so large that the problem becomes computationally impractical.

\begin{figure*}
\centering
\includegraphics[width=\linewidth]{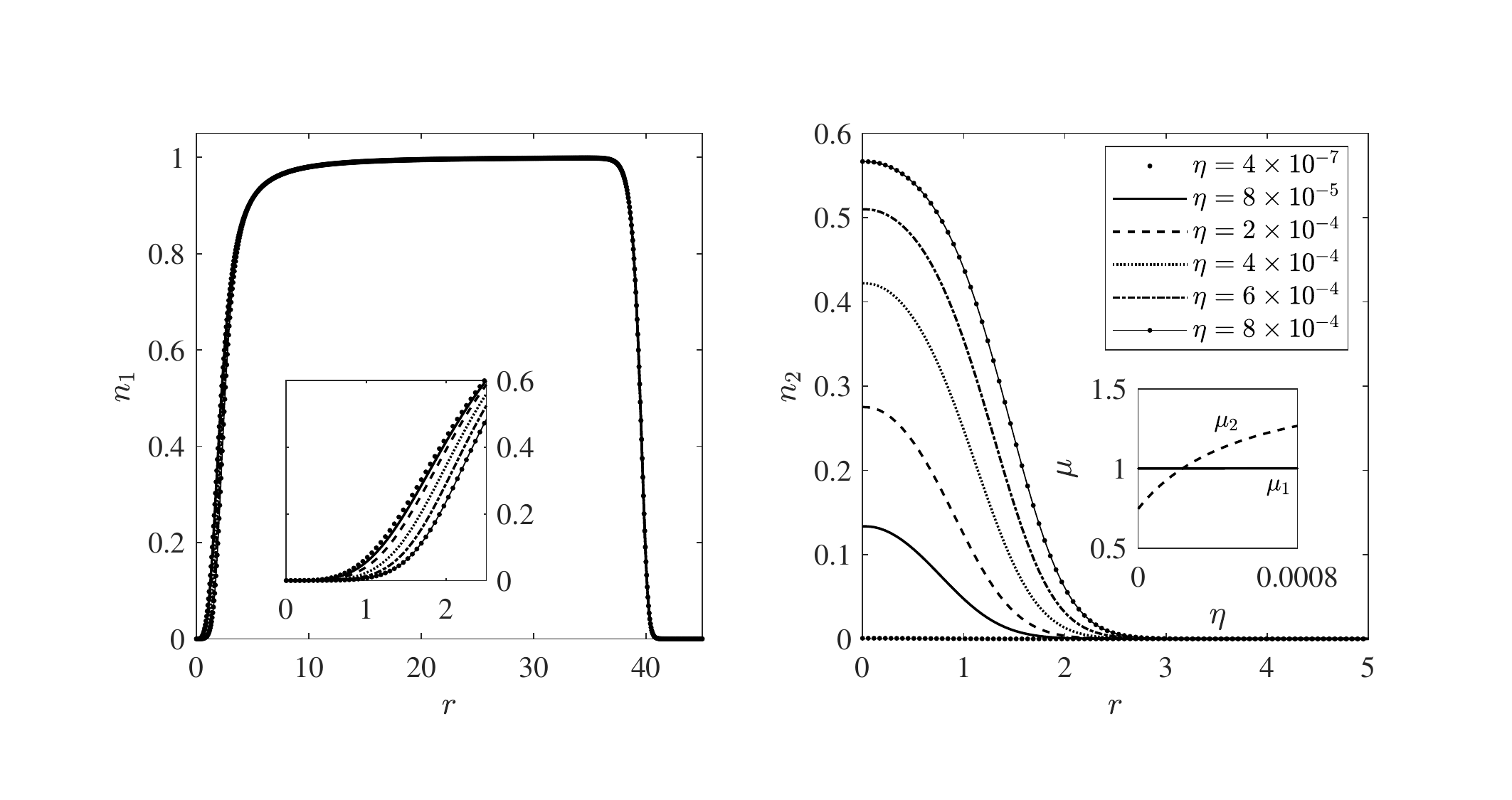}
\caption{Density profiles for species 1 (left) and species 2 (right) for various $\eta$ values (indicated in the figure legend on the right). The vortex axis is $r=0$ is on the left and the trap's boundary is at $r=r_B=40$.
The other parameters are $\ell=2$, and $N_1\simeq 4824$ (the dimensionless number of atoms of species 1 when $\eta=0$).
The inset on the left panel shows a close up of the vortex axis. 
Component 2 occupies the vortex core where the density of component 1 goes to zero.
As the proportion of component 2 increases, the vortex core region widens.
The inset on the right panel shows how the chemical potentials vary with $\eta$.
} \label{fig:dens}
\end{figure*}

\subsection{Stationary solutions}  \label{sec:stat}

To solve for the stationary profiles $n_{1,2}$, we evolve \eqref{GPE2} in imaginary time $\tau=it$.
This amounts to making the replacement $i\partial_t\to-\partial_\tau$ in \eqref{GPE}.
Starting from a sensible initial condition, the population of the higher energy modes will then decrease exponentially with increasing $\tau$, leaving essentially only the ground state at late times.
Since the anticipated densities are independent of the azimuthal coordinates, the speed of the computation is increased by working in polar coordinates.
We therefore solve the following coupled set of equations,
\begin{equation} \label{imag_time}
\begin{split}
    \partial_\tau Y_1 = & \ \left(\frac{\nabla_r^2}{2} -\frac{\ell^2}{2r^2} - V(r)+\mu_1 -|Y_1|^2 -g_{12}|Y_2|^2 \right)Y_1, \\
    \partial_\tau Y_2 = & \ \left(\frac{\nabla_r^2}{2m_2} - V(r)+\mu_2-g_2|Y_2|^2 -g_{12}|Y_1|^2\right)Y_2,
\end{split}
\end{equation}
where $\nabla_r^2 = \partial_r^2+\frac{1}{r}\partial_r$ and $Y_j=|\Psi_j|=\sqrt{n_j}$.
We work on a radial grid $r=[\Delta r,r_B+5]$ with $\Delta r=0.05625$ such that are $800$ radial points and use an imaginary time step $\Delta\tau= 4\times 10^{-4}$.
The evolution is performed using a $4^\mathrm{th}$ order Runge-Kutta algorithm with the radial derivatives approximated using finite difference stencils.

The first step is to fix a value for $N_1$ against which we can compare the effect of increasing $\eta$.
This value is determined by solving \eqref{imag_time} for $\eta=0$, using the fact that $\mu_1(\eta=0)=1$, and integrating over the resulting profile to get $N_1=\int d^2\mathbf{x}n_1(\eta=0)$.
For $r_B=40$ and $\ell=2$, we have $N_1\simeq 4824$. Again, since this is the atom number in dimensionless variables, one must multiply by the dimensionless ratio $d_1/\sqrt{8\pi}a_1$ to find the true total number of atoms.

The second step is to choose an initial condition when the second component is added.
We start by considering case the where no vortex is present ($\ell=0$) with component 2 localised in the centre of the system and component 1 surrounding it.
A subtlety of the problem is that if we want to keep $N_{1,2}$ fixed, the chemical potentials must vary to accommodate these constraints.
A priori, we do not know the values of $\mu_{1,2}$ and must determine them during the imaginary time evolution.
We start by guessing their values using the Thomas-Fermi approximation \eqref{TF}, which in dimensionless units gives $\mu_1=N_1/\pi(r_B^2-r_0^2)$ and $\mu_2=\eta g_2 N_1/\pi r_0^2$.
Since we consider $\eta$ to be very small, we take 1 as our initial guess for the value of $\mu_1$ and invert the first relation to find the value of $r_0$.
The second relation provides a guess for $\mu_2$.
The densities are then $n_1(r_0<r<r_B) = \mu_1$ and $n_2(r>r_0)=\mu_2/g_2$.
We evolve these profiles forward in imaginary time, renormalising $\Psi_{1,2}$ after each time step to keep the values of $N_{1,2}$ fixed.
% Even then, we find that their is a lower energy state available to the system when component 2 sits near the wall of the potential at $r=r_B$. We therefore modify the potential for component 2 so that the wall sits at $r=0.9\,r_B$, ensuring that component stays at the centre. Eventually, we will find that component 2 stays confined to the vortex cores which stay away from the edge of the condensate, hence this modification does not influence our results.
After a long time, the system will have relaxed to its lowest energy state, however, there will be residual time dependence due to the fact that the chemical potentials were chosen incorrectly.
We can use this time dependence to infer the shift in $\mu_j$ using $Y_j(\tau+\Delta\tau)=e^{-\delta\mu_j\Delta\tau}Y_j(\tau)$, so that the true chemical potentials are $\mu_1=1+\delta\mu_1$ and $\mu_2=\eta g_2N_1/\pi r_0^2+\delta\mu_2$ respectively.
The evolution is terminated once the changes $\delta\mu_1$ and $\delta\mu_2$ over a time step are both less than $10^{-6}$.
This is suitably small given that $\mu_j\sim\mathcal{O}(1)$ for both species and is computationally feasible when solving for a large range of $\eta$ values.

In the third step, we imprint a vortex by setting $\ell\neq 0$ in \eqref{imag_time}, taking the ground states obtained from step 2 as the initial condition.
We evolve again in imaginary time (using the same termination criterion) to find the stationary vortex solutions and update the chemical potentials accordingly.

In Fig.~\ref{fig:dens}, we present the resulting density profiles $n_{1,2}$ as a function of $r$ for a vortex with $\ell=2$.
We see that component 2 stays inside the core of the vortex in component 1, and its effect is to widen the core.
It is remarkable that even the small values of $\eta$ which we consider here have such a significant effect.
This widening becomes more significant as $\eta$ increases.
We also display the variation of the chemical potentials with $\eta$ in the inset of the right panel.
$\mu_1$ increases with $\eta$ by a very small amount due to the repulsive effect of species 2 inside the vortex core.
This forces more of the $N_1$ particles into the bulk of the condensate, increasing the mean value of the density in this region.
$\mu_2$ increases much more obviously with $\eta$ since $N_2$ is increasing and therefore the density of species 2 must go up accordingly.
Note however that $\mu_2$ does not vanish as $\eta\to 0$.
This is because there is a contribution of the curvature of $n_2$ to the chemical potential.
Indeed, solving the equation in \eqref{imag_time} for $Y_2$ in limit $r\to 0$ gives the relation $\mu_2 = (g_2 n_2 - n_2''/2m_2n_2)|_{r=0}$.
This does not go to zero as $n_2\to 0$ since the ratio $n_2''/n_2$ stays finite in this limit.

\begin{figure*} 
\centering
\includegraphics[width=\linewidth]{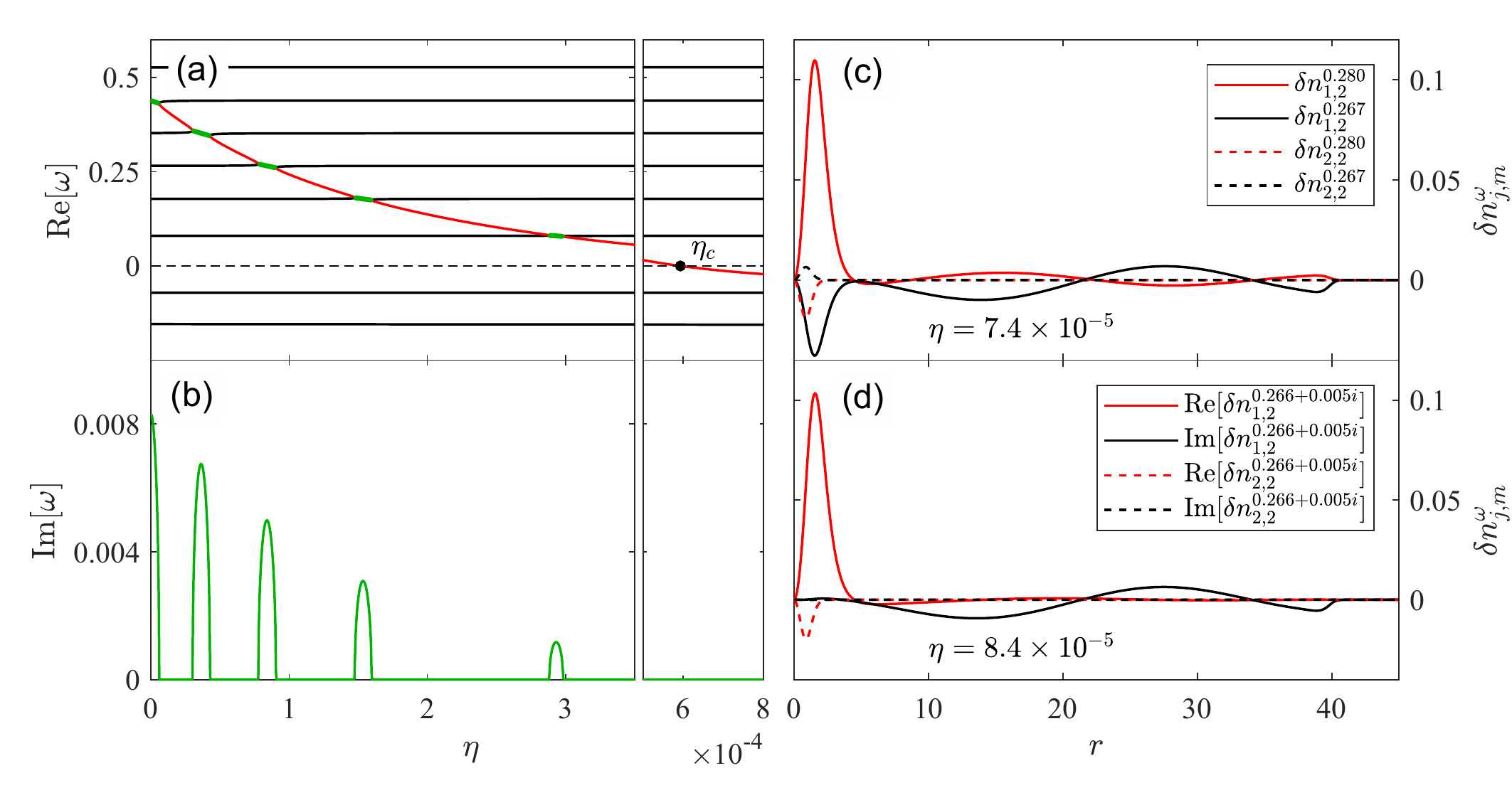}
\caption{On the left we display the eigenvalues $\omega$ of \eqref{BdG} as a function of $\eta$ for $m=\ell=2$ and $r_B=40$, with the real part in panel (a) and the imaginary part in panel (b). 
For clarity, both panels (a) and (b) are split in two regions for small and large values of $\eta$.
The modes whose frequency is approximately independent of $\eta$ (solid black horizontal lines) are bulk excitations (phonons) which are not affected by component 2 in the vortex core.
The mode whose frequency decreases with $\eta$ (solid red) is an excitation of the vortex core and has negative norm for all frequencies.
When the vortex mode crosses a bulk excitation mode, the two modes couple with each other to produce a pair of complex conjugate modes (solid green) with zero norm and complex frequency $\omega$.
At a critical value $\eta_c$ (marked as a solid black dot), the frequency of the vortex mode (red line) changes sign.
On the right, we display some example mode functions for the vortex mode.
In panel (c), we pick an $\eta$ value slightly to the left of the third instability bubble on panel (b).
The solid black (resp. red) curve corresponds to the bulk (resp. vortex) excitations of component 1 with the frequency indicated.
The dashed curves of the same colour correspond to component 2.
In panel (d), we pick the $\eta$ value at the peak of the third instability bubble and display the real and imaginary parts of the growing mode.
Comparing panels (c) and (d), we see that the vortex mode enters mainly the real component of the unstable eigenmode whilst the bulk excitation is captured by the imaginary part.
} \label{fig:freq}
\end{figure*}

\begin{figure} 
\centering
\includegraphics[width=\linewidth]{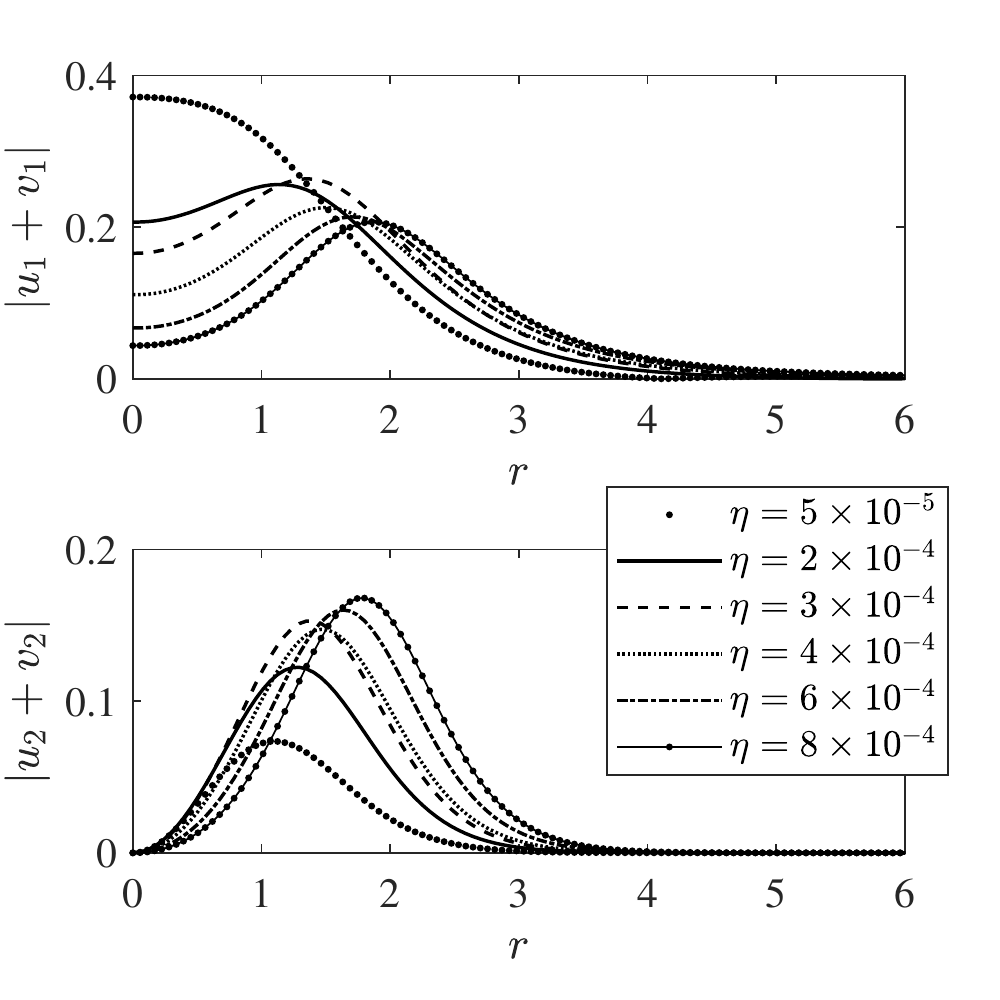}
\caption{Variation of the vortex mode waveform with $\eta$. 
In particular we display the modulus of $u_j+v_j$, which is the combination appearing in formula \eqref{delta_n} for density fluctuations.
The main effect of the second component is to reduce the amplitude of excitations of species 1 on the vortex axis. 
The shift in the maximum of $|u_1+v_1|$ between $\eta=2\times10^{-4}$ and $3\times10^{-4}$ is responsible for $\mathcal{H}_{\mathrm{BdG},1}$ becoming positive on Fig.~\ref{fig:linenergy}, since the waveform moves into a region of positive energy density.
} \label{fig:waveform}
\end{figure}

\begin{figure} %[ht!] 
\centering
\includegraphics[width=\linewidth]{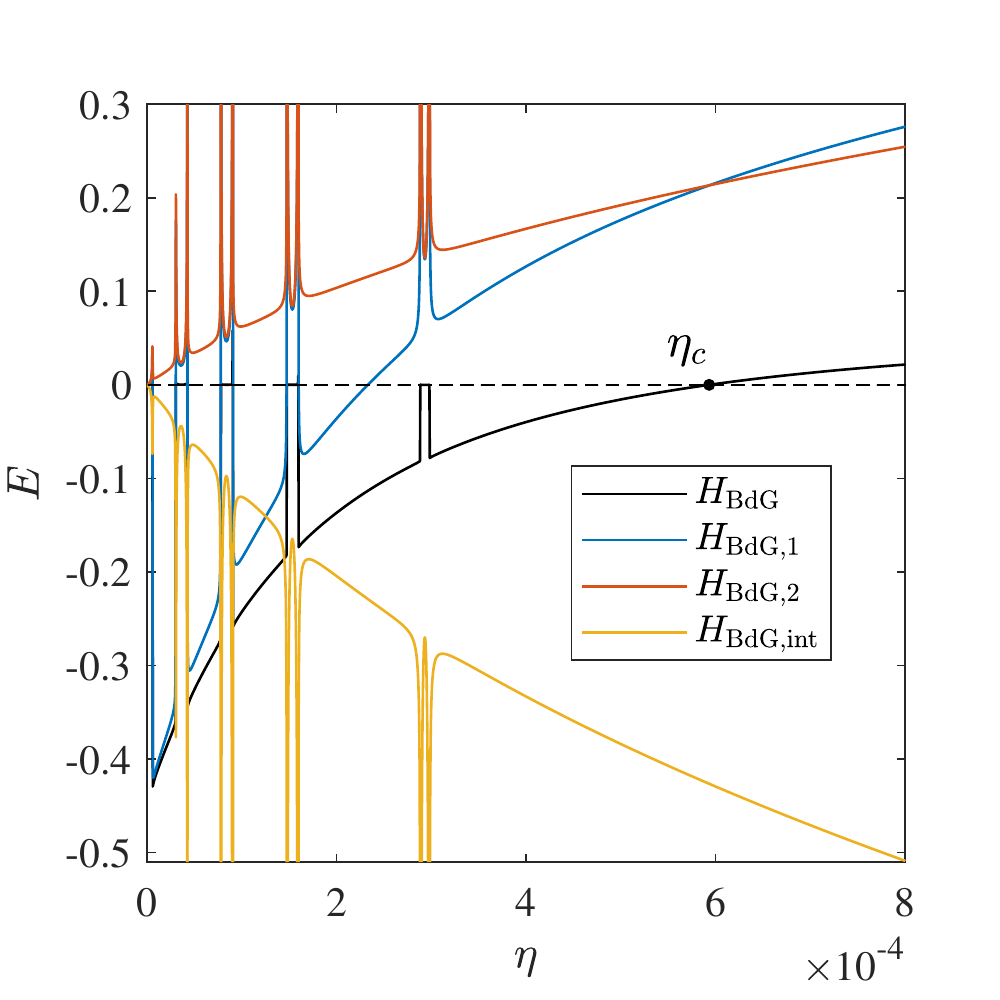}
\caption{Vortex mode energy densities in \eqref{SBdG} integrated over space.
The total energy smoothly increases from negative to positive, except in finite windows where it vanishes due to the dynamical instability.
The change in sign of the total energy is associated with the energy of fluctuations in species 1 exceeding that of species 2.
} \label{fig:linenergy}
\end{figure}

\section{Linear analysis}

Having found the stationary vortex solutions to \eqref{GPE2}, we determine their stability by analysing the normal modes of the system.
It is well-known that MQVs in one-component condensates spontaneously decay into clusters of SQVs and that this instability can be attributed to an unstable mode of the linearised equations of motion \cite{shin2004dynamical}.
We now outline how this analysis can be extended to two-component condensates and demonstrate that, for sufficiently high values of $\eta$, this mode can be stabilised.

\subsection{Bogoliubov method} \label{sec:lin}

Let $\delta\psi_j$ be small fluctuations about $\Psi_j$ and write the following Ansatz,
\begin{equation}
\begin{bmatrix}
\delta\psi_1 \\ \delta\psi_2 \\ \delta\psi_1^* \\ \delta\psi_2^*
\end{bmatrix} = \sum_{m=-\infty}^\infty e^{im\theta} \begin{bmatrix}
u_1e^{i\ell\theta} \\ u_2 \\ v_1e^{-i\ell\theta} \\ v_2
\end{bmatrix},
\end{equation}
with $u_j$ and $v_j$ functions of $r,t$ and $m$, where $m$ is the azimuthal wavenumber. When the equations of motion (\ref{GPE2}) are linearized, since neither $n_{1,2}$ nor $\mathbf{v}_{1,2}$ depend on $\theta$, each $m$ component evolves independently of the others.
Due to the symmetries $u_j^*(m)=v_j(-m)$ and $v_j^*(m)=u_j(-m)$, we can restrict our attention to $m\geq0$ modes without loss of generality.
The fluctuations $|U\rangle=(u_1,u_2,v_1,v_2)^\mathrm{T}$ obey the following coupled set of equations,
\begin{equation} \label{BdG}
i\partial_t |U\rangle = \widehat{L} |U\rangle.
\end{equation}
%with $|U\rangle = (u_1,u_2,v_1,v_2)^\mathrm{T}$. 
The operator $\widehat{L}$ is,
\begin{equation} \label{BdG_matrix}
\widehat{L} = \begin{bmatrix}
D^+_1 & \varepsilon & n_1 & \varepsilon \\
\varepsilon & D_2 & \varepsilon & g_2n_2 \\
-n_1 & -\varepsilon & -D^-_1 & -\varepsilon \\
-\varepsilon & -g_2n_2 & -\varepsilon & -D_2
\end{bmatrix},
\end{equation}
where,
\begin{equation}
\begin{split}
D^\pm_1 = & \ -\frac{1}{2}\left[\nabla_r^2 - \frac{(\ell\pm m)^2}{r^2}\right] + V + 2n_1 + g_{12}n_2 - \mu_1, \\
D_2 = & \ -\frac{1}{2M_2}\left[\nabla_r^2 - \frac{m^2}{r^2}\right] + V + 2g_2n_2 + g_{12}n_1 - \mu_2,
\end{split}
\end{equation}
and the coupling between fluctuations of both species is determined by $\varepsilon=g_{12}\sqrt{n_1n_2}$.
Note that for immiscible fluids, the region of overlap of the two components is small.
$\varepsilon$ will be small in this region and zero everywhere else.
The fields $u_j$ and $v_j$ are related to the components of the density field by,
\begin{equation} \label{delta_n}
    \delta n_{j,m} = \sqrt{n_j}\left(u_j+v_j\right)
\end{equation}
The equations in \eqref{BdG} can be derived from the action $S_\mathrm{BdG} = \int dt d^2\mathbf{x}\mathcal{L}_\mathrm{BdG}$ where,
\begin{equation} \label{SBdG}
\begin{split}
    \mathcal{L}_\mathrm{BdG} = & \ iu_1^*\dot{u}_1-iv_1^*\dot{v}_1 + iu_2^*\dot{u}_2-iv_2^*\dot{v}_2 - \mathcal{H}_\mathrm{BdG}, \\
    \mathcal{H}_\mathrm{BdG} = & \ \mathcal{H}_{\mathrm{BdG},1} + \mathcal{H}_{\mathrm{BdG},2} + \mathcal{H}_{\mathrm{BdG,int}}, \\
    \mathcal{H}_{\mathrm{BdG},1} = & \ u_1^*D_1^+u_1 + v_1^*D_1^-v_1 + n_1(u_1^*v_1+u_1v_1^*), \\
    \mathcal{H}_{\mathrm{BdG},2} = & \ u_2^*D_2u_2 + v_2^*D_2v_2 + g_2n_2(u_2^*v_2+u_2v_2^*) \\
    \mathcal{H}_{\mathrm{BdG,int}} = & \ \varepsilon[(u_1^*+v_1^*)(u_2+v_2) + \mathrm{c.c.}].
\end{split}
\end{equation}
$S_\mathrm{BdG}$ is invariant when each of the fields $u_j$ and $v_j$ simultaneously undergo the same phase rotation.
The corresponding conserved quantity is the Bogoliubov norm,
\begin{equation}
    \mathcal{N}_\mathrm{BdG} = \langle U|\gamma^0|U\rangle \equiv \sum_j\int d^2\mathbf{x}(|u_j|^2-|v_j|^2),
\end{equation}
where $\gamma^0=\mathrm{diag}(1,1,-1,-1)$ is the zeroth gamma matrix.
Notice that due to the presence of the cross-terms with coefficient $\varepsilon$ in \eqref{SBdG}, excitations of species 1 can be converted into excitations of species 2 and vice versa.
However, the total number of excitations, as measured by $\mathcal{N}_\mathrm{BdG}$, is conserved.

Since the background is stationary, we can further decompose the fluctuations into the separate frequency components,
\begin{equation}
    |U(r,t)\rangle = \int^\infty_{-\infty}\frac{d\omega}{2\pi}a_\omega(t)|\widetilde{U}(r;\omega)\rangle,
\end{equation}
where $a_\omega$ and $|\widetilde{U}\rangle$ respectively satisfy,
\begin{equation} \label{BdG2}
    i\partial_t a_\omega = \omega a_\omega, \qquad \omega|\widetilde{U}\rangle = \widehat{L}|\widetilde{U}\rangle.
\end{equation}
The first equation is solved by $a_\omega\propto e^{-i\omega t}$ up to an arbitrary constant, whilst the second must be solved numerically. 
The Hamiltonian for stationary modes is related to the Bogoliubov norm via $H_\mathrm{BdG}=\int d^2\mathbf{x}\mathcal{H}_\mathrm{BdG}=\omega\mathcal{N}_\mathrm{BdG}$.
Therefore a positive frequency mode with negative norm (or a negative frequency mode with positive norm) has negative energy.
The sign of the energy plays an important role in the discussion of the instability.

To solve the second equation in \eqref{BdG2} computationally, we write the radial derivative terms using finite difference stencils on a discrete $r$-grid then find the eigenvalues and eigenfunctions of the resulting matrix $\widehat{L}$.
The eigenfunctions are normalised so that $|\mathrm{Re}[\mathcal{N}_\mathrm{BdG}]|=1$, which gives an unambiguous normalisation condition for all solutions obtained.
We place a Dirichlet boundary condition at large $r$, although this does not play a significant role as it is applied in a region where $V(r)$ is large, hence, modes with energy lower than the barrier height quickly decay to zero inside this region.
A discussion of the boundary conditions at $r=0$ can be found in Appendix~\ref{app:BCs}.

In Fig.~\ref{fig:freq}, we present the eigenvalues $\omega$ as a function of $\eta$ for $\ell=2$ and $r_B=40$.
There are two distinct types of excitation.
Bulk excitations (shown in black) predominantly occupy the region outside of the vortex core and are associated with collective excitations of the condensate, or phonons at low wavenumbers.
Indeed, on panel (c), it can be seen that these modes have a large oscillatory component in the large $r$ region.
They have positive energy ($H_\mathrm{BdG}>0$) and are largely unaffected by increasing $\eta$, which influences only the vortex core.
There is a second class of excitation which we refer to as vortex modes.
These have negative energy ($H_\mathrm{BdG}<0$) at low $\eta$  and are mainly localised around the vortex axis, as illustrated in panel (c).
In panel (a), we see that there is only one such mode for the $\ell=2$ vortex represented by the red line.
Since this mode occupies the same region as component 2, it is strongly affected when $\eta$ is increased. 
Indeed, the frequency of the vortex mode on panel (a) can be seen to decrease rapidly even for relatively small amounts of component 2.
In the presence of a dissipation mechanism, these modes will be excited since they lower the total energy of the system, i.e. they are an energetic instability.

When $\mathrm{Re}[\omega]$ of the vortex mode crosses that of a bulk excitation, the two couple to produce a complex conjugate pair of modes with vanishing Bogoliubov norm.
This leads to the appearance of isolated bubbles in $\mathrm{Im}[\omega]$ in panel (b).
The mode with $\mathrm{Im}[\omega]>0$ represents a dynamical instability, that is, the central vortex spontaneously sheds its energy into a sound wave in the bulk even without a dissipation mechanism.
This does not violate energy conservation since the total energy of the mode encoding this behaviour is zero ($H_\mathrm{BdG}=0$).
The complex conjugate mode describes the time reversed process, i.e. absorption of a sound wave by the central vortex.
In panel (d), we show how the waveforms of the two modes in panel (c) combine to produce the real and imaginary parts of the instability.

Note that for large enough $r_B$ (larger than considered here) i.e. in the limit that finite size effects become negligible, the bubbles in $\mathrm{Im}[\omega]$ eventually merge into a continuum \cite{giacomelli2020ergoregion}.
In that case, there is a dynamical instability at $\eta=0$ for all larger trap sizes and, in particular, in the infinite system limit.

We also show in Fig.~\ref{fig:waveform} the effect of increasing $\eta$ on the waveform of the vortex mode around the vortex axis.
The main feature is that the amplitude of the mode at $r=0$ is suppressed as the number of atoms in species 2 is increased.
This will be addressed in the next section.

For large $\eta$, the vortex mode has a frequency lower than all positive frequency bulk excitations.
At a critical $\eta$ value, which we call $\eta_c$ (the black dot in Fig.~\ref{fig:freq}(a)), the frequency of the vortex mode changes sign.
Since the Bogoliubov norm of this mode is negative, the energy of the vortex mode is positive for $\eta>\eta_c$ and it will no longer get excited when a dissipation mechanism is included.
Furthermore, since the negative frequency bulk excitations have negative norm, they have positive energy and the vortex mode cannot mix with these modes to produce a dynamical instability with zero energy.
Above $\eta_c$, the system is stabilised by the presence of the second component.
The next section will explore the mechanism underpinning this stabilisation whilst Section~\ref{sec:num} will demonstrate stability via a full numerical simulation of \eqref{GPE2}.

From Fig.~\ref{fig:freq}(a), we find the value $\eta_c=5.933\times 10^{-4}$.
Since the value obtained will in general depend on the resolution used in the numerics, we estimate that the infinite resolution limit would yield the value $\eta_c=5.936\times 10^{-4}$ (details in Appendix~\ref{app:res}).
Note that this value is also specific to an $\ell=2$ vortex in a trap of a radius $r_B=40$ and will in general depend on both these parameters.
Due to computational cost, we do not study this dependence here.

A computation of the mode energy in \eqref{SBdG} gives insight into the behaviour around $\eta_c$.
In Fig.~\ref{fig:linenergy}, we integrate the (vortex mode) energy density components in \eqref{SBdG} over space and plot as a function of $\eta$.
The total energy $H_\mathrm{BdG}$ is negative at low $\eta$ except inside finite windows where it is vanishing due to the dynamical instability.
At the edge of these windows, the amplitudes of the normalised mode diverges, as reflected in the components of $H_\mathrm{BdG}$, but does so in a way that keeps the total energy finite.
The energy of species 2 fluctuations $H_{\mathrm{BdG},2}$ is always positive whilst the interaction energy $H_\mathrm{BdG,int}$ between fluctuations of each species is always negative.
The energy of species 1 fluctuations $H_{\mathrm{BdG},1}$ is initially negative but becomes positive around $\eta=2.33\times10^{-4}$.
Correspondingly, the maximum amplitude of the vortex mode waveform shifts outwards into a region where it has positive energy density (see Fig.~\ref{fig:waveform}).
Since $H_\mathrm{BdG,int}$ is close to the value of $-2H_{\mathrm{BdG},2}$ the total mode energy is still negative even when $H_{\mathrm{BdG},1}=0$.
Interestingly, in this regime, the energetic instability of the vortex mode is a result of the interaction between fluctuations of the two species, rather than being an inherent property of the vortex in species 1 alone.
Increasing $\eta$ further, the total energy changes sign at $\eta_c$. This coincides with $H_{\mathrm{BdG},1}$ exceeding the value of $H_{\mathrm{BdG},2}$.

\begin{figure}
\centering
\includegraphics[width=\linewidth]{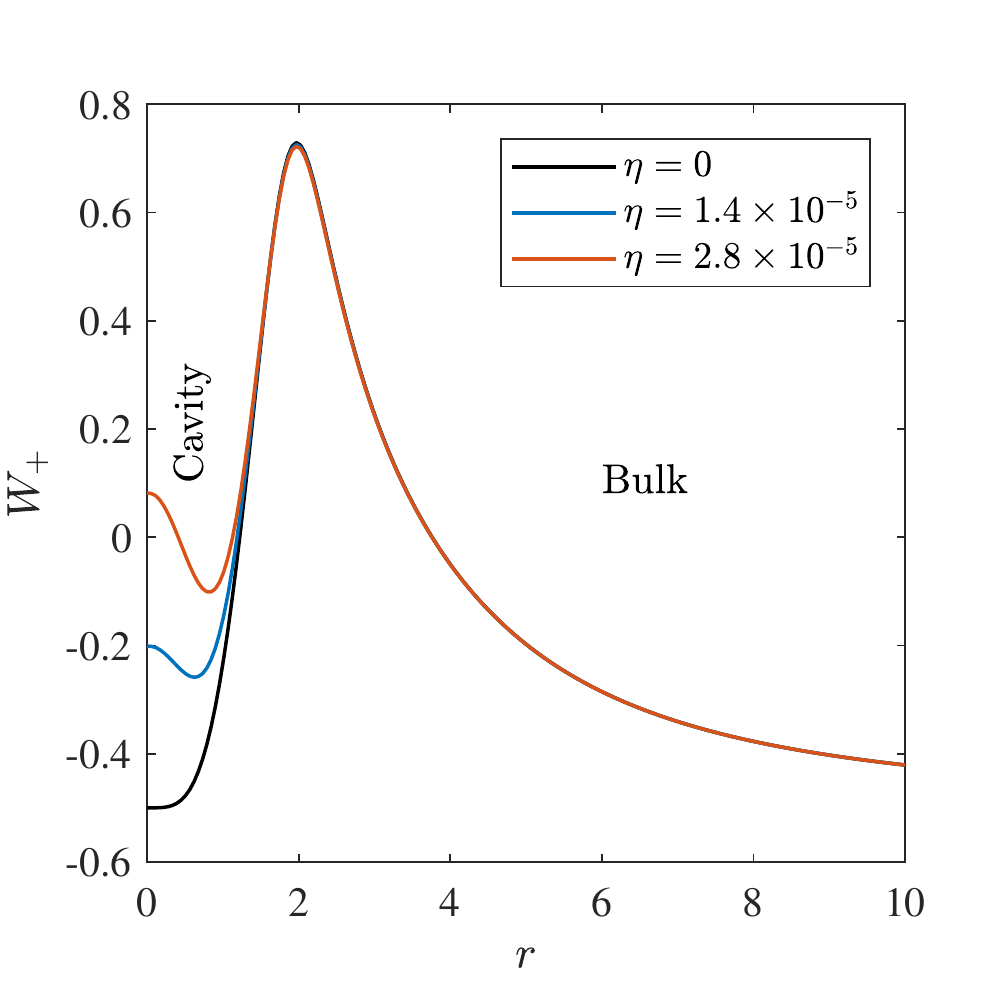}
\caption{The potential $W_+$ from \eqref{W_eff} with $m=\ell=2$ for an example frequency $\omega=0.75$.
In the WKB approximation, excitations of component 1 propagate when $W_+<0$.
Excitations of the vortex are bound states inside the cavity centred on the axis and correspond to the central vortex splitting into a cluster.
A second component in the vortex core induces a bulge in the cavity, decreasing the amount of space for vortex excitations.
As the bulge grows in size, the vortex excitation decreases its frequency until, for large enough $\eta$, there are no positive frequency modes which fit inside the cavity.
} \label{fig:potential}
\end{figure}

\subsection{WKB method}

\begin{figure*} 
\centering
\includegraphics[width=\linewidth]{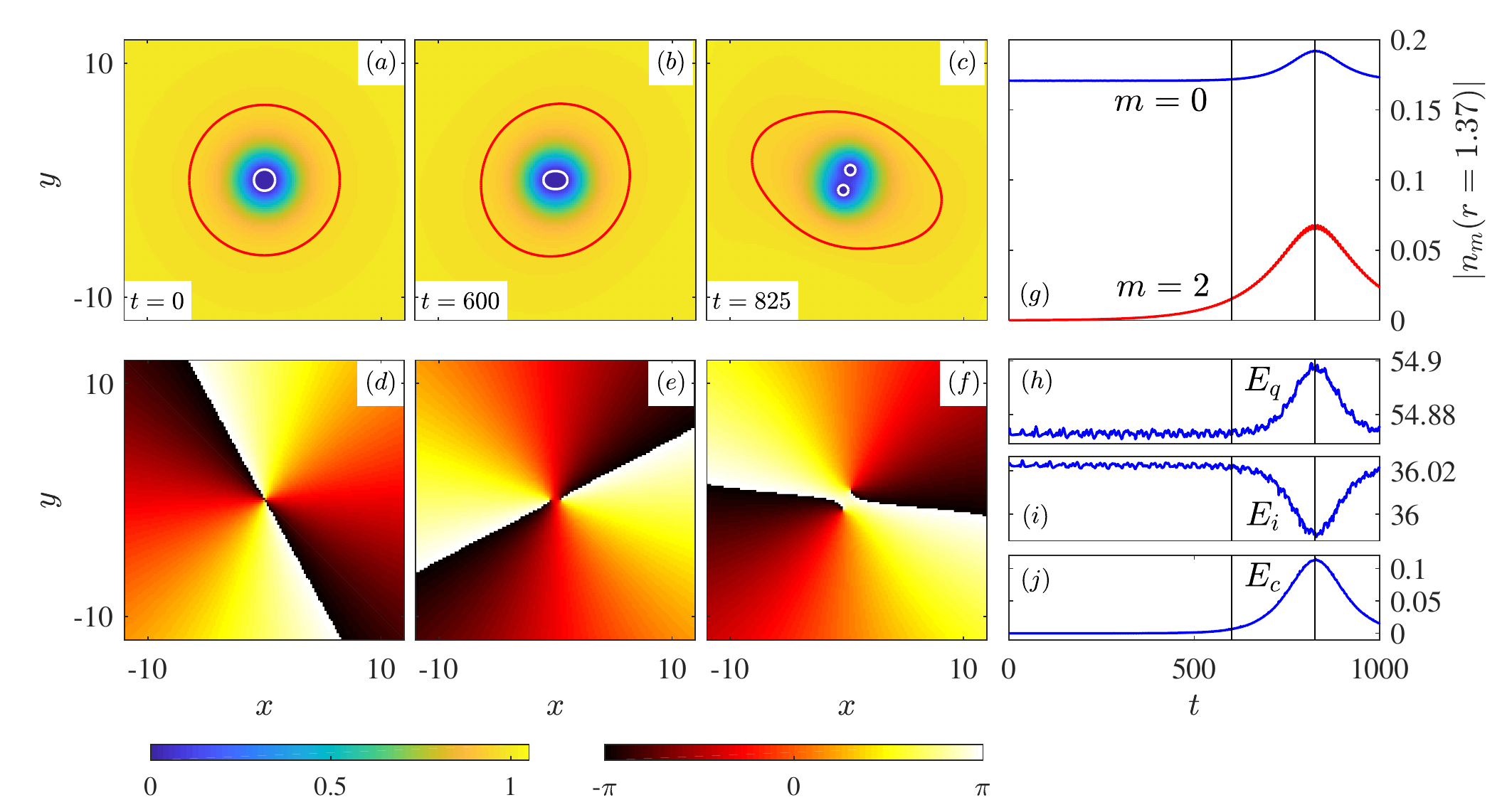}
\caption{Splitting of a vortex with $\ell=2$ when $\eta=0$. The density is displayed on panels (a-c) at three different times. 
The constant density contours are set at $0.95$ (red outer contour) and $0.05$ (white inner contours). These help indicate the vortex splitting and the production of an $m=2$ sound wave in the bulk of the condensate.
Panels (d-f) indicate the phase at the same times.
In panel (g), we show the variation of the modulus of the $m=0$ and $m=2$ Fourier components of the density with time.
The $m=2$ mode grows exponentially before saturating and decaying due to finite size effects \cite{patrick2022origin}.
The vertical black lines indicate the times $t=600,825$.
We display various components of the energy in panels (h-j), showing that the vortex splitting is associated with a decrease (increase) in incompressible (compressible/quantum) energy.
} \label{fig:onecomp}
\end{figure*}

To interpret why the vortex mode is shifted to lower frequency with increasing $\eta$, we apply a WKB method.
This method has been demonstrated to give insight into the physical origin of vortex instabilities \cite{patrick2022origin} as well as providing reasonable estimates of the unstable mode frequencies \cite{patrick2022quantum}.
Following the procedure outlined in \cite{patrick2022quantum}, we first write a WKB Ansatz,
\begin{equation}
\begin{bmatrix}
u_j \\ v_j
\end{bmatrix} = \begin{bmatrix}
\mathcal{A}_j(r) \\ \mathcal{B}_j(r)
\end{bmatrix}e^{i\int p(r) dr},
\end{equation}
where $p$ is the local component of the wavevector in the radial direction and we assume $|p^2|\gg|\partial_r p|$, $|p\mathcal{A}_j|\gg|\partial_r \mathcal{A}_j|$ and similarly for $\mathcal{B}_j$.
Taking only the leading order contributions and eliminating the four amplitudes, \eqref{BdG2} becomes,
\begin{equation} \label{disps}
\mathcal{D}_1(k_1)\mathcal{D}_2(k_2) = \varepsilon^2 k_1^2k^2_2,
\end{equation}
where,
\begin{equation}
\begin{split}
\mathcal{D}_1(k_1) = & \ \Omega^2 - n_1k_1^2 - k_1^4/4,  \\
\mathcal{D}_2(k_2) = & \ \omega^2 - g_2n_2k_2^2 - k_2^4/4,
\end{split}
\end{equation}
and $\Omega=\omega-m\ell/r^2$ is the frequency in the frame rotating with the vortex.
We have also defined a pair of effective wavevectors by,
\begin{equation}
    k_1^2 = p^2 + \frac{\m_1^2}{r^2}, \qquad M_2k_2^2 = p^2 + \frac{\m_2^2}{r^2}
\end{equation}
where the effective azimuthal numbers are,
\begin{equation}
\begin{split}
\m^2_1 = & \ m^2+\ell^2 + 2r^2(n_1+g_{12}n_2-\mu_1), \\
\m^2_2 = & \ m^2 + 2M_2r^2(g_2n_2+g_{12}n_1-\mu_2).
\end{split}
\end{equation}
Hence, we can identify the condition in \eqref{disps} as two coupled dispersion relations respectively describing excitations of components 1 and 2 interacting through a small overlap region where $\varepsilon$ is non-zero.
As a first approximation, we study the non-interacting case with $\varepsilon\to 0$.
We are particularly interested in the fluctuations of component 1 (since this is where the instability arises) which approximately satisfy $\mathcal{D}_1(k_1)\simeq 0$.
This condition has two sets of solutions for $p$ which are decoupled from each other. 
The first type are evanescent over the whole system whereas the second can propagate.
We are only interested in the second type.
These are determined by,
\begin{equation}
p^2 + W_+ = 0, \qquad W_+ = -2\sqrt{n_1^2+\Omega^2} + 2n_1 + \frac{\m_1^2}{r^2}.
\end{equation}
Points where $p=0$ correspond to turning points, i.e. radial locations where an incoming wave instantaneously comes to rest before reversing its direction and propagating back out \cite{patrick2022quantum}.
We can therefore view $W_+$ as an effective potential; when $W_+<0$ these waves are propagating whilst for $W_+>0$ they are evanescent. 
%Solving $W_+=0$ for $\omega$ defines two curves,
%\begin{equation}
%\omega_\pm = \frac{m\ell}{r^2}\pm\sqrt{n_1\frac{\m^2_1}{r^2}+\frac{\m^4_1}{4r^4}},
%\end{equation}
%such that when $\omega=\omega_+$ (or $\omega=\omega_-$) there is a turning point. Frequencies above $\omega_+$ have positive Bogoliubov norm whereas those below $\omega_-$ have negative norm. Between the two curves, modes are evanescent.

To see the effect of adding a second component, one can expand the $\m_1^2$ term in $W_+$:
\begin{equation} \label{W_eff}
W_+ = -2\sqrt{n_1^2+\Omega^2} + 4n_1-2\mu_1 + \frac{m^2+\ell^2}{r^2} + 2g_{12}n_2.
\end{equation}
In the absence of the second component, this potential has a cavity located about $r=0$ where waves with correct frequency can become trapped as bound states
These states approximately satisfy $\cos I\simeq 0$ \cite{patrick2022quantum}, where $I=\int_{_\mathrm{cav}}\!p\,dr$ is the phase integral inside the cavity in the vortex core (see Appendix \ref{app:I} for a further discussion).
When a second species is present inside the vortex core, the final term in \eqref{W_eff} creates a bulge in the potential around $r=0$ (as shown in Fig.~\ref{fig:potential}) thereby decreasing the amount of space the mode has to fit in the cavity.
This bulge inhibits wave propagation in the immediate vicinity of the vortex axis, explaining the feature of Fig.~\ref{fig:waveform} where the amplitude of the waveform drops on the vortex axis with increasing $\eta$.
Conventional wisdom gained from the 1D Schr\"odinger equation in quantum mechanics would dictate that this should raise the frequency of the bound state, however, this is not the case here as we now illustrate.

To solve the resonance condition in the cavity, one should compute $I(\omega)$ for all frequencies which probe the cavity before finding the one which solves $\cos I=0$.
For small $\eta$, we show in Appendix~\ref{app:I} that the integral near the bottom of the cavity (for $m=\ell=2$) can be expanded as,
\begin{equation} \label{Iaprx}
    I(\omega) = \frac{\pi(\omega_0-\omega)}{|2g_{12}n''_{2,0}|^{1/2}} + \mathcal{O}[(\omega_0-\omega)^2],
\end{equation}
where $\omega_0$ is the maximum frequency which probes the cavity and $n''_{2,0}$ is the curvature of the density of component 2 on the axis.
This expression demonstrates that $I$ decreases as $\omega$ increases, i.e. $\partial_\omega I<0$.
More generally, the reason for this difference with the standard Schr\"odinger picture (in which $\partial_\omega I>0$) is that the vortex mode in the WKB picture is on the lower branch of the dispersion relation where $\Omega<0$ (which is related to the negative norm of the mode).
On this branch, the root with $p>0$ has negative group velocity $\partial_p\omega<0$, i.e. decreasing $\omega$ increases the value of $p$.
Therefore, since increasing $\eta$ will decrease $p$ through the final term in \eqref{W_eff}, the frequency of the bound state decreases to keep $I$ fixed at the value satisfying the resonance condition.

Although the WKB method correctly predicts the decrease of the vortex mode frequency with increasing $\omega$, it fails at providing the quantitative value. This is to be expected since the approximation works best for high frequencies in regions where the background functions change gradually.
Indeed, in \cite{patrick2022quantum}, the WKB estimates are practically indistinguishable from numerical calculations for bulk excitations, whilst for the vortex mode (which occupies a region where the background density quickly changes) the WKB estimate differs notably from the numerics.
When $\eta\neq0$, the frequency is lower than the $\eta=0$ case and the bulge in $W_+$ makes background variations more significant.
Both of these features decrease the quantitative power of the WKB approximation.

\section{Numerics} \label{sec:num}

Having gained an understanding of the instability mechanism from the linear theory, we now perform a series of fully non-linear simulations of \eqref{GPE2}.
The first goal will be to demonstrate explicitly the stabilisation mechanism, showing that for high $\eta$ values the $\ell=2$ vortex fails to split.
We then move on to discuss a by-product of the stabilisation mechanism: the counter-rotating orbital motion of a pair of co-rotating vortices.

\begin{figure*}
\centering
\includegraphics[width=\linewidth]{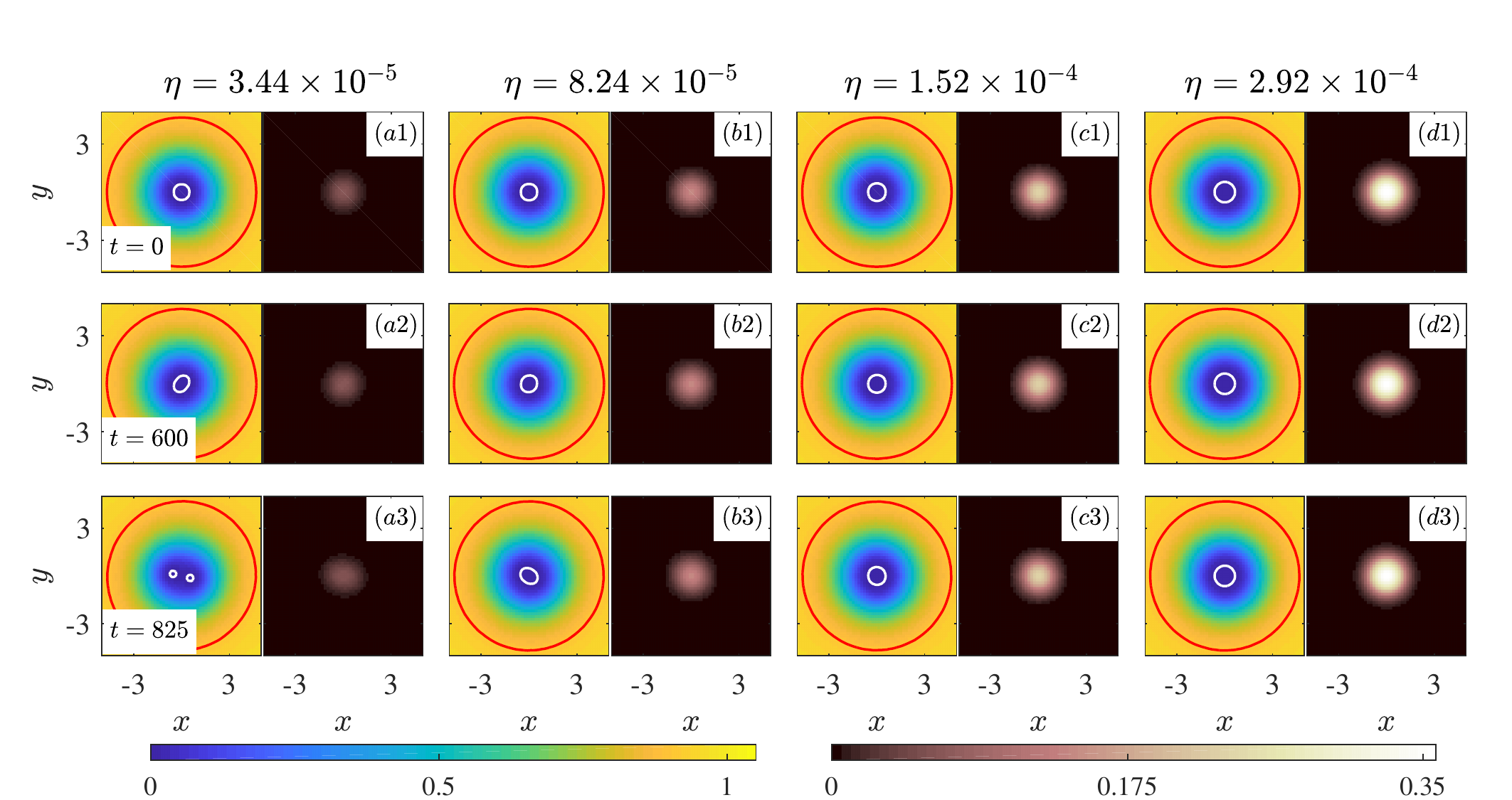}
\caption{The densities $n_{1,2}=|\Psi_{1,2}|^2$ for various $\eta$ at the same times shown in Fig.~\ref{fig:dens}.
On each panel, the density of component 1 is shown on the left (corresponding colorbar bottom left) and component 2 on the right (bottom right)
The constant density contours for component 1 are set at $0.9$ (red outer contour) and $0.005$ (white inner contours) and help indicate the vortex splitting.
As $\eta$ is increased, the central vortex in component 1 is less deformed at a given time, whilst the density of component 2 increases, becoming more and more visible on the colorscale.
} \label{fig:dens2}
\end{figure*}

\subsection{One component}

We first evolve \eqref{GPE} for $\eta=0$, i.e. when the second species is absent.
For our spatial grid, we take $x\in[-(r_B+5),r_B+5]$ and similarly for $y$ with $N=512$ spatial points.
We take $\Delta t=5\times 10^{-3}$ and evolve using a split-step Fourier method for $2\times 10^5$ steps in time, saving every $200^\mathrm{th}$ frame.
For the initial condition, we seed the stationary vortex profile for $\ell=2$ and $r_B=40$ with the unstable mode of amplitude $|a_\omega(t=0)|=10^{-3}$.
We note that the corresponding resolution in the radial direction is $256$ points, which differs from the value of $800$ used in Section~\ref{sec:stat}.
The reduced resolution slightly shifts the location of the instability bubbles in Fig.~\ref{fig:freq}.
Hence, for consistency, we repeat the procedure in Section~\ref{sec:stat} and \ref{sec:lin} for $\Delta\tau=5\times 10^{-3}$ and 256 radial grid points when producing the initial condition, since we want to precisely seed the mode which is unstable in our non-linear simulations.
A comparison of the normal mode spectra at different resolutions can be found in Appendix~\ref{app:res}.

The results are shown in Fig.~\ref{fig:onecomp}.
Panels (a-c) show the density at early, intermediate and late times, with constant density contour superimposed in red/white, and panels (d-f) show the phase at the same times.
In panels (a) and (d), the vortex is close to its stationary profile whilst in panels (b) and (e), a deformation of the central vortex can be seen.
In panels (c) and (f), the central vortex has split into two $\ell=1$ vortices.
The density contours indicate that this splitting is accompanied by the production of a bulk excitation, namely, a sound wave with $m=2$.
At sufficiently late times, the vortices move back into the centre due to finite system effects studied in \cite{patrick2022origin}.
Since our focus is on the splitting mechanism, we do not discuss this behaviour further here. In the presence of dissipation, the initial splitting described here continues and the vortices separate widely, as we show in Section~\ref{sec:damp}.

In panel (g), we perform a Fourier decomposition of the density $n=(2\pi)^{-1}\int d\theta e^{im\theta}n_m$ and display modulus of the Fourier components $m=0,2$ at the location $r=1.37$. This location corresponds to the peak of the radial waveform $n_{m=2}(r)$.
At first, the $m=2$ mode grows exponentially in line with the prediction of Fig.~\ref{fig:freq}.
Around $t=800$, the $m=2$ mode reaches a maximum amplitude and starts to decay as the vortices move back into the centre.
Near the maximum amplitude of $n_{m=2}$, we also observe the non-linear effect of the unstable mode on the $m=0$ component.
This is simply the backreaction of the growing mode onto the background density profile, which can be physically understood as the condensate filling in the centre as the vortices move apart.

The instability is associated with a decrease of incompressible (vortex) energy and an increase in compressible (sound wave) energy.
The energy of the system can be decomposed as,
\begin{equation}
\begin{split}
    & E = E_\mathrm{kin} + E_q + E_\mathrm{pot} + E_\mathrm{int}, \quad E_a = \int d^2\mathbf{x}\,\mathcal{E}_a \\
    & \mathcal{E}_\mathrm{kin} = \tfrac{1}{2}n|\grad\Phi|^2, \qquad \mathcal{E}_q = \tfrac{1}{2}|\grad\sqrt{n}|^2,
\end{split}
\end{equation}
where the kinetic energy $E_\mathrm{kin}$ is associated with a non-vanishing velocity field and the quantum energy $E_q$ results from density variations.
$\mathcal{E}_\mathrm{pot}=Vn$ and $\mathcal{E}_\mathrm{int}=n^2/2$ are the usual potential and interactions energy densities respectively.
The kinetic energy can be further split into compressible and incompressible parts by defining $\mathbf{u}=\sqrt{n}\grad\Phi$ and writing $\mathbf{u}=\mathbf{u}_c+\mathbf{u}_i$, where the incompressible part obeys $\grad\cdot\mathbf{u}_i=0$.
Operationally, the compressible part can be found by performing a Fourier transform $\mathcal{F}$, projecting $\mathbf{u}$ onto the wavevector $\mathbf{k}$ and computing the inverse Fourier transform $\mathcal{F}^{-1}$. That is, $\mathbf{u}_c = \mathcal{F}^{-1}[\mathbf{k}(\mathbf{k}\cdot\mathcal{F}\mathbf{u})/||\mathbf{k}||^2]$ and $\mathbf{u}_i=\mathbf{u}-\mathbf{u}_c$.
Since $\mathbf{u}_c$ and $\mathbf{u}_i$ have zero overlap, the kinetic energy can be written $E_\mathrm{kin}=E_c+E_i$, with $E_c=\int d^2\mathbf{x} \,\mathbf{u}_c$ and similarly for $E_i$.
We display $E_q$, $E_i$ and $E_c$ in panels (h), (i) and (j) respectively.
The incompressible energy decreases as the vortices move apart (into a lower energy configuration) and the compressible and quantum energies increase in response. 
The two vertical black lines in the rightmost panels correspond to the times in panels (b) and (c) respectively.

\begin{figure}
\centering
\includegraphics[width=\linewidth]{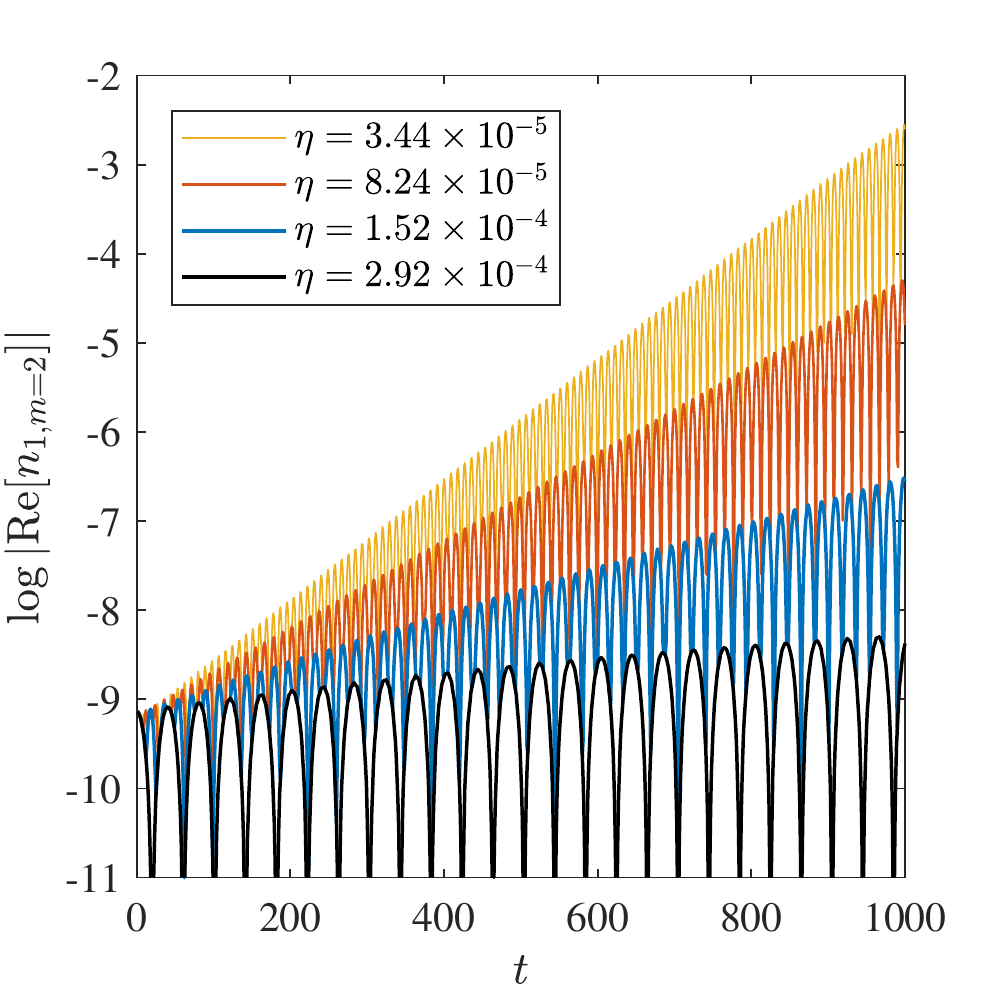}
\caption{Growth of the $m=2$ unstable mode for various $\eta$ values.
For each $\eta$, we look at the time dependence of the mode at the spatial location where $n_{1,m=2}$ has maximum amplitude (which is around $r\sim 1.6-1.9$).
As $\eta$ increases, the oscillation frequency and growth rate decreases in accordance with the linear prediction of Fig.~\ref{fig:freq}.
} \label{fig:growth}
\end{figure}

\subsection{Two components}

To assess the effect of adding a small amount of component 2 inside the vortex core of component 1, we repeat the simulation of the last section for various values of $\eta$.
In particular, we start from the stationary profiles for condensate 1 and 2 and perturb the system by inserting the unstable mode with an amplitude $10^{-3}$.
The values of $\eta$ chosen correspond to the peaks of $\mathrm{Im}[\omega]$ in the stability windows at the chosen resolution.

In Fig.~\ref{fig:dens2}, we display the densities of species 1 and species 2 for four values of $\eta$ at the same times depicted as the $\eta=0$ case in Fig.~\ref{fig:dens}.
As $\eta$ increases, we clearly see that the vortices have separated less at a given instant of time when compared with the single component system, thereby demonstrating the stabilising effect of the second species.
Since the minute separation becomes less and less transparent from the density plots with increasing $\eta$, we show on Fig.~\ref{fig:growth} the growth of the $m=2$ mode in the vortex core.
Here, we clearly see that both the oscillation frequency and growth rate decrease with increasing $\eta$.
The values of $\mathrm{Re}[\omega]$ and $\mathrm{Im}[\omega]$ correspond to those predicted by the linear analysis in Fig.~\ref{fig:freq} (see Appendix~\ref{app:res} for an explicit comparison).

\begin{figure*} 
\centering
\includegraphics[width=\linewidth]{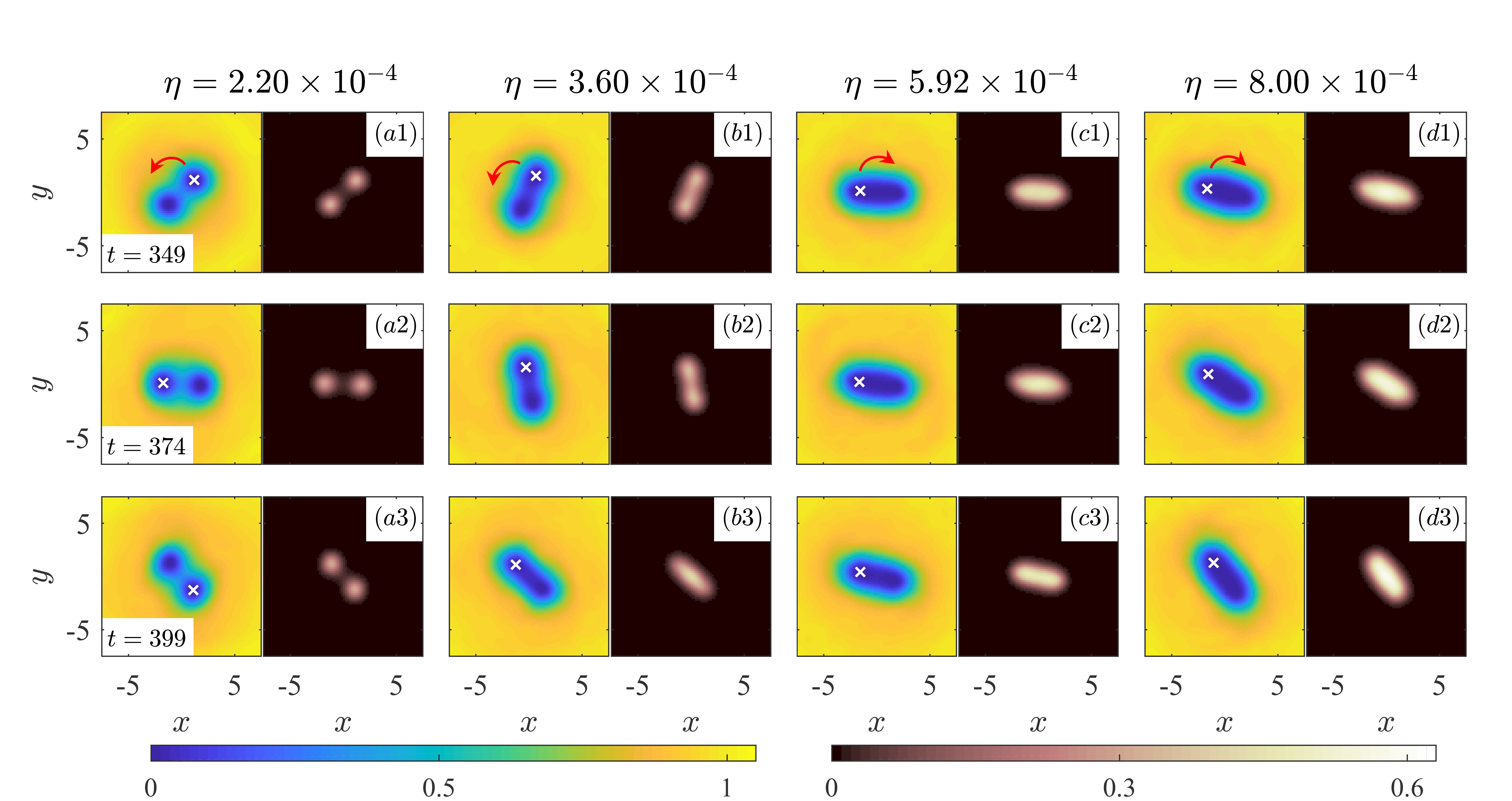}
\caption{
Vortex pair motion for various $\eta$ values at three different times.
In each of the four columns, the left plot depicts $|\Psi_1|^2$ (colorbar bottom left) and the right plot $|\Psi_2|^2$ (colorbar bottom right).
Each row corresponds to the time indicated on the left.
The direction of the vortices is indicated with a red arrow and one of the vortices is marked with  a white cross to help illustrate the motion.
In panels (a1-3) the pair complete less than half a full orbit in the counter-clockwise sense between each snapshot.
In panels (b1-3), the motion is also counter-clockwise with lower orbital frequency.
In panels (c1-3), the motion is clockwise with low orbital frequency and in panels (d1-3) the motion is clockwise and slightly faster.
The change in direction of the orbit is associated with a low density region between the vortex pair in component 1 and a high density in the same region in component 2.
} \label{fig:reverse}
\end{figure*}

\subsection{Orbit reversal}

In general, the unstable mode in the spectrum of the MQV is associated with its splitting into a cluster of SQVs, whose distribution is determined by the $m$-fold symmetry of the unstable mode.
In single component systems, the cluster rotates in the direction of the net winding of the system, since each vortex essentially moves under the collective influence of all the others.
The story is not so simple when a second component is added.
The orbital frequency of the resulting cluster is given by $\mathrm{Re}[\omega]/m$, with $\mathrm{Re}[\omega]$ the oscillation frequency of the unstable mode.
Hence, the frequency inversion for $\eta > \eta_c$ seen in Fig.~\ref{fig:freq} is associated with the reversal of the cluster's orbital motion.
We demonstrate this now with a numerical simulation.

We choose a series of $\eta$ values where the vortex is dynamically stable.
Note here that dynamical stability below $\eta_c$ results from finite size effects for certain ranges of $r_B$, whilst above $\eta_c$ it persists for all trap sizes, as shown by Fig.~\ref{fig:freq}.
Using the same resolution as the previous section, we now seed the stationary vortex profile with the vortex mode of amplitude $|a_\omega(t=0)|=2$.
Due to the large amplitude, in all cases we observe a sudden burst of sound from the central vortex as the system relaxes through non-linear interactions.
After this initial burst, we observe a stably orbiting vortex pair in the centre of the trap surrounded by small noise.
On Fig.~\ref{fig:reverse}, we display snapshots of the density profiles of the two species at three different times.
The red arrows indicate the direction of the orbital motion.
For the two leftmost columns, the orbital motion is counter-clockwise, corresponding to $\mathrm{Re}[\omega]>0$ for the vortex mode below $\eta_c$.
The atoms of species 2 are shared between the two $\ell=1$ vortices in the pair and track their motion.
Conversely, the orbital motion is clockwise on the two rightmost columns, corresponding to $\mathrm{Re}[\omega]<0$ above $\eta_c$.

For large $\eta$ we find that a region of depleted density in component 1 persists between the two separated vortices and is occupied by component 2.
We interpret this as the result of the waveforms in Fig.~\ref{fig:waveform}.
For $\eta=0$, the maximum value of the waveform on the vortex axis non-linearly excites the $m=0$ mode as the vortices separate, causing atoms to fill in the space between them.
For large $\eta$, the waveform decreases approaching $r=0$, suppressing the non-linear excitation of $m=0$ in that region.
Physically, the presence of component 2 on the vortex axis prevents excitations in component 1 propagating there, which means there is nothing to increase the value of the density of component 1 at $r=0$ as the vortices separate.

\subsection{Dissipation} \label{sec:damp}

\begin{figure*} 
\centering
\includegraphics[width=\linewidth]{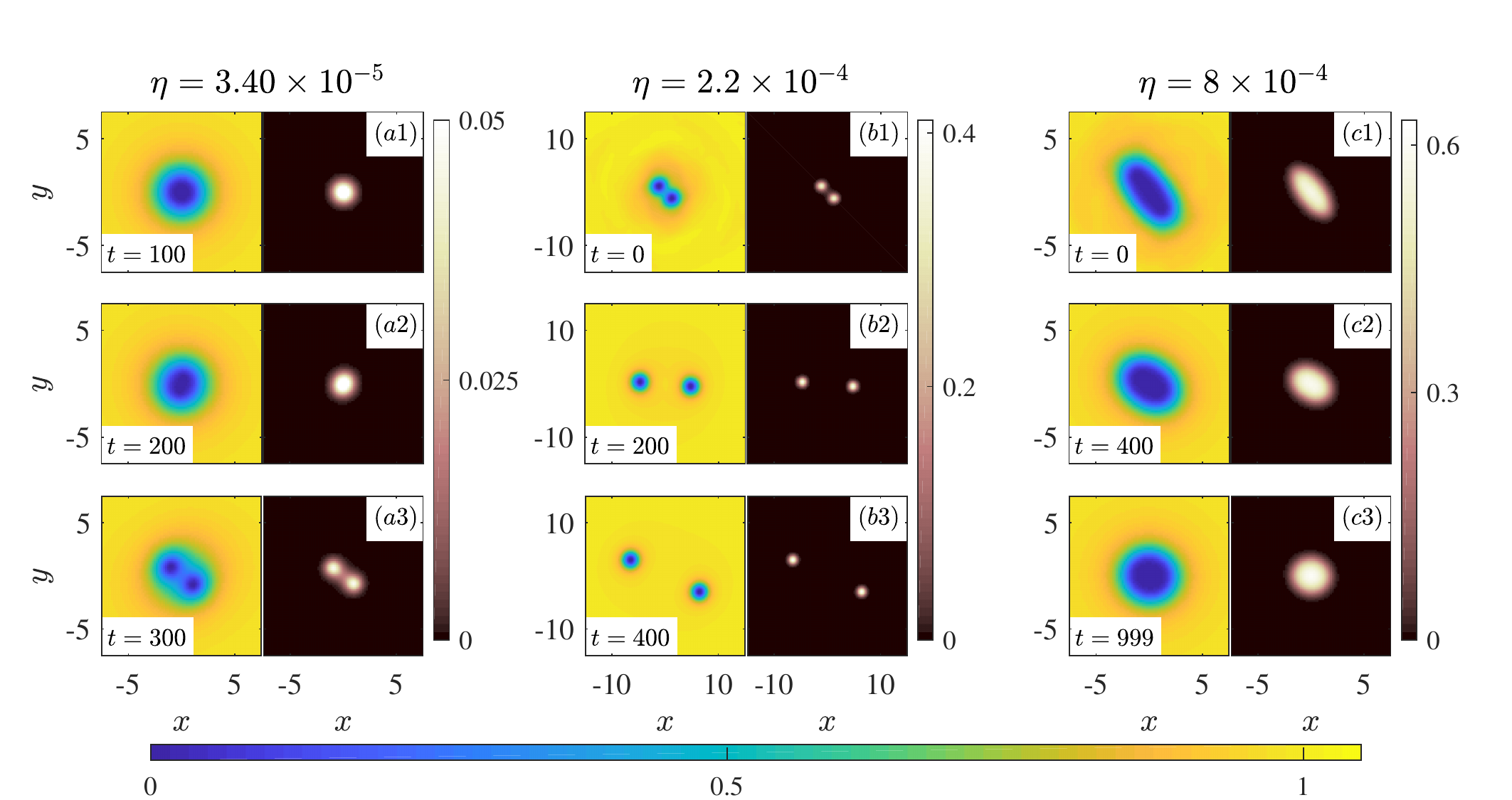}
\caption{Vortex pair dynamics with damping for various $\eta$.
In panels (a1-3), atoms of the second species stay confined to the SQV vortex cores when the MQV decays.
In panels (b1-3) the vortices spiral out in an $\eta$ independent manner, with the species 2 atoms tracking the motion of the vortex cores.
In panels (c1-3), $\eta$ is above the critical fraction, hence, the system lowers its energy by recombining into a single, central MQV.
} \label{fig:damp}
\end{figure*}

In the presence of a dissipation mechanism, an MQV in a one-component condensate will still decay even if the vortex mode does not couple to any bulk excitations.
The reason is that the vortex mode has negative energy, hence, its excitation lowers the total energy of the system, as required by the dissipation.
The same is not necessarily true in a two-component system as we now show, since the vortex mode has a positive energy above $\eta_c$.

In Fig.~\ref{fig:damp}, we present results of three simulations performed with damping.
The initial conditions in panels (a), (b) and (c) are, respectively, the final frames of the simulations shown in Fig.~\ref{fig:dens2}(a), Fig.~\ref{fig:reverse}(a) and Fig.~\ref{fig:reverse}(d), corresponding to the same $\eta$ values indicated at the top of the figures.
These respectively correspond to the lowest fraction of $\eta$ simulated which are unstable (panel (a)) and stable (panel (b)) when $\gamma_1=0$ and the largest $\eta$ value simulated above $\eta_c$ (panel (c)).
As $\eta$ is small, damping applied to atoms of the second species will quickly reduce $N_2$ to zero.
Hence, we only apply damping to the first component.
This is achieved by replacing $i\partial_t$ by $(i-\gamma_1)\partial_t$ in \eqref{GPE2}, taking the value $\gamma_1=5\times 10^{-2}$.

In Fig.~\ref{fig:damp}(a1-3), the initial condition is a dynamically unstable vortex mode with low amplitude. In the three snapshots, we observe the central MQV splitting into a pair of SQVs rotating counter-clockwise with equal amounts of second component in the two cores.
During the splitting, the second species remains confined to the vortex cores, rather than being dispersed across the system.

In panels (b1-3), the initial condition is two nearby SQVs surrounded by noise.
The noise is quickly damped out and the vortices spiral out in the counter-clockwise sense with decreasing orbital frequency.
In this regime, the vortices orbit in accordance with the regular point-vortex dynamics \cite{barenghi2016primer} which dictates the orbital frequency is approximately $\Omega=2/s^2$, where $s$ is the separation between vortices, with the atoms of species 2 playing a negligible role in the dynamics.
In accordance with this expectation, the dynamics observed for $t>300$ at $\eta=3.4\times10^{-5}$ (i.e. after the last frame shown on panel (a)) also follow the point-vortex behaviour.
From this, we conclude that the modified dynamics in two component condensates is absent when species 2 atoms are split into distinct parcels which occupy the separate vortex cores.
In other words, the stabilisation is a feature of a single parcel of species 2 atoms shared across the vortices.

To support this conclusion, in panels (c1-3), we take such a configuration as our initial condition.
The corresponding vortex mode at this $\eta$ value has positive energy, hence, the separation of vortices (whilst retaining a single parcel of species 2 atoms) is associated with an energy gain relative to the stationary MQV.
As the system evolves, the vortices orbit in the clockwise sense and damping reduces the vortex separation to lower the energy of the system.
At late times, we recover a stationary, central MQV.
This demonstrates that MQVs seeded with species 2 atoms above a critical fraction can be inherently stable objects even when a dissipation mechanism is present.

\section{Discussion}

The results presented here show that, in a two-component condensate confined in a bucket trap in the immiscible regime, multiply-quantised vortices (MQVs) in the first component can be stabilised when a relatively small number of atoms of the second component collect in the vortex core.
We have found that if the fraction $\eta$ of atoms of the second component is small, the growth rate of the dynamically unstable mode which causes vortex splitting is smaller than when the second component is absent.
We found there exists a critical fraction $\eta_c$ above which vortex decay is energetically unfavorable.
Hence, for $\eta>\eta_c$, MQVs are stabilised even in the presence of a dissipation mechanism since it would cost energy to separate them.
A consequence of the stabilisation mechanism is that nearby co-rotating vortices can orbit in the opposite sense to their individual rotations when enough atoms of the second component are shared across the cores.

The parameters which we have used in this model were chosen such that an experiment would be reasonably feasible. 
The greatest challenge in performing this experiment would be creating a trapping potential profile that directly matches what we have used here; however, since the first realisation of a uniform trapping potential \cite{navon2016emergence}, advancements have been made in generating homogeneous condensates \cite{johnstone2019evolution,gauthier2019giant}, with traps of the size of $\approx100\mathrm{\mu m}$.
We also expect this phenomenon to persist in harmonic traps since the instability is a feature of the vortex core rather than the details of the trapping potential (see e.g. \cite{giacomelli2020ergoregion} for an example where the normal mode spectra of MQVs in harmonic and box-like trapping potentials are qualitatively the same in one component condensates).

Our results stand in contrast to those in \cite{berti2022superradiant}, where a miscible, two-component condensate (with equal mass and self-interaction parameters) was shown to be more unstable than vortices in one component condensates. This is a result of the region where spin waves have negative energy density extending outside the vortex core.
It is an interesting parallel then that in the opposite case, where the mixtures are immiscible, MQVs can be made more or completely stable.

Our results further demonstrate that even in dynamic scenarios, the atoms of the second component remain attached to the vortex cores rather than becoming scattered about the trap.
Therefore the proposed method of MQV stabilisation could provide a solution to the problem of destructive imaging. 
The most common method to image a condensate is absorption imaging, which consists of illuminating the cold gas with a laser and measuring the shadow cast by absorption \cite{anderson1999spatial}.
Although this method provides a detailed image of the condensate, irradiating the system drastically changes its energy and dynamics, and in the worst case destroys the condensate \cite{pyragius2012developing}.
It is also possible to remove a small fraction of the atoms of a specific component of a condensate from a trap and to image them independently \cite{serafini2015dynamics}. 
If one stabilised vortex cores with a second component, the use of a specific microwave pulse could outcouple only secondary component atoms, resulting in an image describing vortex core locations without affecting the dynamics of the primary component. As this stabilisation has proven robust, vortex cores undergoing rapid motion could be reliably imaged and tracked.

An added consequence of the stabilisation is that MQVs (and also SQVs) are endowed with an extra degree of freedom relative to one component condensates where an excited vortex can only radiate sound.
Here, the extra degree of freedom is the excitations of the second component which would allow for more complicated interactions in systems containing more than one vortex.
This situation has some analogy with superfluid $^3$He, where vortices may have a double vortex core whose rotational symmetry about the vortex axis is broken \cite{thuneberg1986}. By creating a natural ribbon, the double vortex structure would help the definition of superfluid helicity \cite{hayder2017}.

Another interesting question relates to the dynamics of vortices in 3D condensates. 
If atoms of the second component are unevenly distributed along a pair of nearby vortex filaments, different segments of the vortex filament could feasibly orbit in opposite directions. 
This may induce vortex reconnections, with potential consequences for superfluid turbulence in two-component mixtures.

A further possibility we raise involves the nucleation of vortices containing atoms of the second component.
In a system of species 1 atoms in which vortices are absent, a small amount of species 2 could be added close to the edge of the trapping potential where the density drops to zero.
Rotating the condensate would lead to vortex nucleation close to the edge of the trap, and it is feasible such vortices would carry atoms of the second species.
Furthermore, as we have seen, for large enough proportions of atoms in species 2, MQVs can be energetically preferable to SQVs, raising the intriguing possibility of nucleating vortices with large winding numbers. 

Finally, since the vortex mode has vanishing oscillation frequency at the critical fraction $\eta_c$, it would be interesting to investigate the non-linear dynamics in the vicinity of this value to see if static (i.e. non-orbiting) configurations of vortices can be constructed.

\section*{Acknowledgments}
We are grateful to Tom Billam and Connor F. Swales for discussions.
For the purpose of open access, the author has applied a Creative Commons Attribution (CC BY) licence to any Author Accepted Manuscript version arising.
This work was supported in part by the STFC Quantum Technology Grants ST/T005858/1 (SP \& RG), ST/T006900/1 (CFB) and Kings College London (AG).
RG also acknowledges support from the Perimeter Institute.
Research at Perimeter Institute is supported by the Government of Canada through the Department of Innovation, Science and Economic Development Canada and by the Province of Ontario through the Ministry of Research, Innovation and Science.

\bibliography{biblio2.bib}
\bibliographystyle{apsrev4-2}

\newpage

\appendix
\section{Dimensional reduction} \label{app:dim}

The action for the two component system in three-dimensional (3D) space is,
\begin{equation*}
    \mathcal{S}^\3 = \int dt d^2\mathbf{x}\left[ \sum_j \left(i\hbar{\Psi^\3_j}^*\dot{\Psi}^\3_j - \mathcal{H}^\3_j\right) - \mathcal{H}^\3_I\right],
\end{equation*}
with,
\begin{equation*}
\begin{split}
    \mathcal{H}^\3_j = & \ {\Psi^\3_j}^*\left[\hat{K}^\3_j+V_j^\3(\mathbf{x})-{\tilde \mu^\3_j}\right]\Psi^\3_j + \frac{G^\3_j}{2}|\Psi^\3_j|^4, \\
    \mathcal{H}_I = & \ G^\3_{12}|\Psi^\3_1|^2|\Psi^\3_2|^2.
\end{split}
\end{equation*}
We now assume a tight confinement in the vertical ($z$) direction and seek an action governing the dimensionally reduced dynamics in the $(x,y)$ plane.
To this end, we write,
\begin{equation*}
    V^\3_j(x,y,z) = V(x,y) + \frac{1}{2}M_j\omega_j^2 z^2, \qquad \tilde{\mu}^\3_j = \tilde{\mu}_j + \frac{\hbar\omega_j}{2}
\end{equation*}
where $V(x,y)$ and $\tilde{\mu}_j$ are the 2D bucket trap and chemical potentials in the main text and the $z$ dependent piece of $V^\3_j$ is a harmonic trap of frequency $\omega_j$.
The break down of the chemical potential in this way amounts to assuming the $z$ dependent piece of the wavefunction is the ground state if we also assume $\tilde{\mu}_j\ll\hbar\omega_j$ and $\Psi_j^\3(x,y,z,t) = \Psi_j^\2(x,y,t)f_j(z)$.
The 3D action can then be split as $\mathcal{S}^\3=\mathcal{S}_z+\mathcal{S}$ where $\mathcal{S}_z\sim\mathcal{O}(\hbar\omega_j)$ and $\mathcal{S}\sim\mathcal{O}(\tilde{\mu}_j)$.
The vertical part is,
\begin{equation*}
\begin{split}
    \mathcal{S}_z = & \ \int dt d^2\mathbf{x} |\Psi_j^\2|^2 \int dz\sum_j\bigg(\frac{\hbar\omega_j}{2}|f_j|^2- \\
    & \qquad \qquad \qquad \frac{\hbar^2}{2M_j}|\partial_z f_j|^2 - \frac{1}{2}M_j\omega_j^2 z^2|f_j|^2\bigg),
\end{split}
\end{equation*}
whose variation yields,
\begin{equation}
    \frac{\hbar^2}{2M_j}\partial^2_z f_j + \left(\frac{\hbar\omega_j}{2}-\frac{1}{2}M_j\omega_j^2z^2\right)f_j = 0,
\end{equation}
which has solutions,
\begin{equation}
    f_j(z) = e^{-z^2/2d_j^2}, \qquad d^2_j = \frac{\hbar}{M_j\omega_j}.
\end{equation}
The $f_j$ can then be inserted into the 2D action and the $z$ integrals evaluated.
Terms involving $|\Psi_j^\2|^2$ (including the derivative term), $|\Psi_j^\2|^4$ and $|\Psi_1^\2\Psi_2^\2|^2$ are proportional to the following $z$ integrals,
\begin{equation}
    \int dz f_j^2 = d_j\sqrt{\pi}, \qquad \int dz f_j^4 = d_j\sqrt{\pi/2},
\end{equation}
and,
\begin{equation}
    \int dz f_1^2 f_2^2 = d_{12}\sqrt{\pi}, \qquad d^{-2}_{12} = d^{-2}_1 + d^{-2}_2.
\end{equation}
Finally, we obtain the action in \eqref{action} once we make the identifications,
\begin{equation}
    \Psi_j = (d_j\sqrt{\pi})^\frac{1}{2}\Psi_j^\2,
\end{equation}
and,
\begin{equation}
    G_j = \frac{G^\3_j}{\sqrt{2\pi}d_j}, \qquad G_{12} = \frac{d_{12}G^\3_{12}}{\sqrt{\pi}d_1 d_2}.
\end{equation}
The interaction parameters are related to the scattering lengths $a_j$ through \cite{pattinson2013equilibrium},
\begin{equation}
    G^\3_j = \frac{4\pi\hbar^2 a_j}{M_j}, \qquad G^\3_{12} = \frac{2\pi\hbar^2 a_{12}}{M_{12}}
\end{equation}
where $M_{12}=M_1M_2/(M_1+M_2)$ is the reduced mass of the two species.
The dimensionless parameters defined in Section~\ref{sec:param} become,
\begin{equation}
    g_2 = \frac{d_1M_1 a_2}{d_2M_2 a_1}, \qquad g_{12} = \frac{d_{12}M_1a_{12}}{\sqrt{2}d_2 M_{12}a_1},
\end{equation}
or in terms of the harmonic trapping frequencies,
\begin{equation}
    g_2 = \sqrt{\frac{M_1\omega_2}{M_2\omega_1}}\frac{a_2}{a_1}, \qquad g_{12} = \frac{1+M_1/M_2}{\sqrt{1+\frac{M_1\omega_1}{M_2\omega_2}}}\frac{a_{12}}{\sqrt{2}a_1}.
\end{equation}
The vertical confinement applied to the two species is the same provided $M_1\omega_1^2=M_2\omega_2^2$.
In this case, we find,
\begin{equation*}
    g_2 = \left(\frac{M_1}{M_2}\right)^{3/4}\frac{a_2}{a_1}, \qquad g_{12} = \frac{1+M_1/M_2}{\sqrt{1+\left(M_1/M_2\right)^{1/2}}}\frac{a_{12}}{\sqrt{2}a_1}.
\end{equation*}
The s–wave scattering lengths for the ${}^{87}$Rb and ${}^{133}$Cs
mixture are $a_1 = 100a_0$, $a_2 = 280a_0$ and $a_{12} = 650a_0$
[26] where $a_0$ is Bohr radius \cite{pattinson2013equilibrium}.
Using $M_1/M_2 = 87/133$, we obtain the values,
\begin{equation}
    g_2 = 2.04, \qquad g_{12} = 5.65,
\end{equation}
which are the values quoted in the main text.

\section{Boundary conditions} \label{app:BCs}

As $r\to 0 $, we have $n_1\to 0$ and $\varepsilon\to 0$.
The fields $u_1$ and $v_1$ each obey the Schr\"odinger equation, whose regular solutions are of the form,
\begin{equation}
u_1 \sim J_{m+\ell}(\alpha_+ r), \qquad v_1 \sim J_{m-\ell}(\alpha_- r)
\end{equation}
with $\alpha_\pm^2 = 2(1-g_{12}n_2\pm\omega)$.
Thus, the boundary conditions are,
\begin{equation}
\begin{split}
m=\ell: & \quad u_1(r=0) = \partial_r v_1|_{r=0} = 0, \\
m=-\ell: & \quad \partial_r u_1|_{r=0} = v_1(r=0) = 0, \\
|m|\neq\ell: & \quad u_1(r=0) = v_1(r=0) = 0.
\end{split}
\end{equation}
The fields $u_2$ and $v_2$ remained coupled to one-another, although they do decouple from $u_1$ and $v_1$.
To find their asymptotic form, we can expand in power series about $r=0$ and apply the method of Frobenius.
Writing,
\begin{equation}
u_2 = \sum_{n=0}^\infty b^+_n r^{n+p^+}, \qquad v_2 = \sum_{n=0}^\infty b^-_n r^{n+p^-}
\end{equation}
the equations of motion become,
\begin{equation}
\begin{split}
& \sum_{n=0}^\infty\left[(n+p^+)^2-m^2\right]b^+_n r^{n+p^+} \\
& + \sum_{n=2}^\infty \left[\beta^+b^+_{n-2}-2g_2n_2 b^-_{n-2}\right]r^{n+p^+} = 0, \\
& \sum_{n=0}^\infty\left[(n+p^-)^2-m^2\right]b^-_n r^{n+p^-} \\
& + \sum_{n=2}^\infty \left[\beta^- b^-_{n-2}-2g_2n_2 b^+_{n-2}\right]r^{n+p^-} = 0,
\end{split}
\end{equation}
with $\beta^\pm = 2(\mu_2-2g_2n_2\pm\omega)$.
The first term in the series yields the values of $p^\pm$. The regular solutions are each given by $p^\pm=|m|$.
Hence the asymptotics at the origin are $u_2\sim r^{|m|}$ and $v_2\sim r^{|m|}$
leading to the boundary conditions,
\begin{equation}
\begin{split}
m=0: & \quad \partial_r u_2|_{r=0} = \partial_r v_2|_{r=0} = 0, \\
m\neq0: & \quad u_2(r=0) = v_2(r=0) = 0.
\end{split}
\end{equation}

\section{Resolution} \label{app:res}

From the procedure outlined in Section~\ref{sec:stat}, we obtain the chemical potentials $\mu^0_j = \mu_j + \Delta\mu_j$, where $\mu_j$ is the true value that one would obtain in the continuum limit.
By checking for a range of resolutions, we observe that the error $\Delta\mu_j$ scales linearly with $\epsilon_\tau=\Delta\tau$ and $\epsilon_r=\Delta r^2$.
This error will offset the normal mode frequencies calculated in Fig.~\ref{fig:freq}, and therefore the value of $\eta_c$ obtained from looking where the vortex mode has zero frequency.

We account for this offset using the following linear perturbation theory argument.
We start with the resolution used to produce Fig.~\ref{fig:freq}, $\epsilon_\tau^0=4\times 10^{-4}$ and $\epsilon_r^0=(45/800)^2$,
and expand the vortex mode frequency in the vicinity of the zero crossing,
\begin{equation}
    \omega = \omega^0 + \partial_\eta\omega^0(\eta-\eta_c^0),
\end{equation}
where $\eta_c^0$ is our prediction for the critical $\eta$ value at this resolution and $\omega^0=0$ by definition.
The gradient $\partial_\eta\omega^0$ is measured from the Fig.~\ref{fig:freq}.
Now consider a change in the resolution.
The expansion above to leading order becomes,
\begin{equation}
    \omega = \omega^0 + \delta\omega + \partial_\eta\omega^0(\eta-\eta_c^0),
\end{equation}
where $\delta\omega$ is the shift in the value of $\omega$ at $\eta_c^0$.
Solving $\omega=0$ gives the new value of $\eta_c$,
\begin{equation}
    \eta_c = \eta_c^0 - \frac{\delta\omega}{\partial_\eta\omega_0}.
\end{equation}
Now, to compute $\eta_c$ at infinite resolution, we need to estimate $\delta\omega$ as we take $\epsilon_\tau$ and $\epsilon_r$ to zero.
To do this, we use the fact that $\omega$ depends linearly on $\epsilon_\tau$ and $\epsilon_r$ and expand the value of $\omega$ at $\eta_c^0$,
\begin{equation}
    \omega(\eta_c^0) = \omega^0_\tau(\epsilon_\tau-\epsilon^0_\tau) + \omega^0_r(\epsilon_r-\epsilon^0_r),
\end{equation}
with,
\begin{equation}
    \omega^0_\tau\equiv\frac{\partial\omega}{\partial\epsilon_\tau}\bigg|_{\epsilon^0_{\tau},\epsilon^0_r}, \qquad \omega^0_r\equiv\frac{\partial\omega}{\partial\epsilon_r}\bigg|_{\epsilon^0_{\tau},\epsilon^0_r}
\end{equation}
We can estimate these gradient terms by independently varying the resolutions away slightly from their initial values.
Once $\omega^0_r$ and $\omega^0_\tau$ are known, the frequency shift at infinite resolution is given by,
\begin{equation}
    \delta\omega = -\omega^0_\tau\epsilon_\tau^0 -\omega^0_r\epsilon_r^0
\end{equation}
The expression for $\eta_c$ in the continuum limit is therefore,
\begin{equation}
    \eta_c = \eta_c^0 + \frac{\omega^0_\tau\epsilon_\tau^0 +\omega^0_r\epsilon_r^0}{\partial_\eta\omega_0}.
\end{equation}
Using $\eta_c^0=5.933\times10^{-4}$, we estimate the true value to be $\eta_c=5.936\times10^{-4}$.

\begin{figure}
\centering
\includegraphics[width=\linewidth]{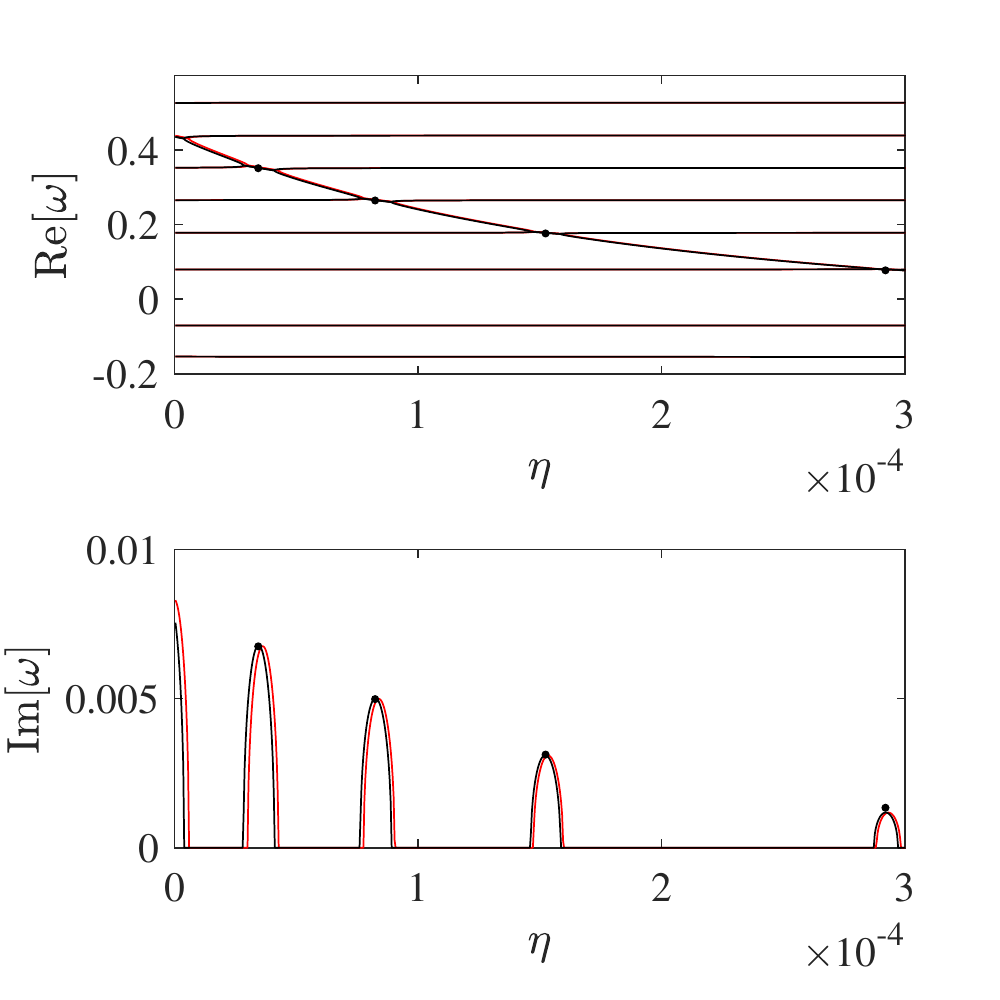}
\caption{Normal mode spectra for resolutions $\Delta\tau=4\times10^{-4}$ and 800 radial points (red) used in Fig.~\ref{fig:freq} and $\Delta\tau=5\times10^{-3}$ and 256 radial points (black) comparable to that used in Section~\ref{sec:num}.
} \label{fig:res}
\end{figure}

When performing the non-linear simulations of Section~\ref{sec:num}, it was not computationally feasible to use the same resolution used for Fig.~\ref{fig:freq}.
In Fig.~\ref{fig:res}, we compare the spectrum for $\Delta\tau=4\times10^{-4}$ and 800 radial points (red curves) with the same for $\Delta\tau=5\times10^{-3}$ and 256 radial points (black lines).
Bulk excitations (horizontal lines) are insensitive to the change since they occupy a region where there are a large number of grid points. The main difference is in the vortex mode frequency, since this mode is covered by far fewer grid points at lower resolution.
This leads to a slight displacement of the instability bubbles in the plot of $\mathrm{Im}[\omega]$.
We also show with black dots the values of $\mathrm{Re}[\omega]$ and $\mathrm{Im}[\omega]$ extracted from the full numerics, namely Fig.~\ref{fig:growth}.
We observe good agreement for all $\eta$ values, although the largest $\eta$ point is displaced slightly above its true value on the plot of $\mathrm{Im}[\omega]$.
We interpret this as being due to finite size effects similar to those studied in \cite{patrick2022origin}, which onset earlier for this particular mode and make it more difficult to extract the growth rate.

\section{Small $r$ integral} \label{app:I}

Here we evaluate the phase integral inside the vortex core in the small $r$ approximation.
The integral is
\begin{equation}
    I = \int^{r_+}_{r_-}\sqrt{-W_+}dr.
\end{equation}
For $W_+(0)>0$, $r_\pm$ are the smallest $r$ values for which $W_+=0$ whilst for $W_+(0)<0 $, $r_-=0$ and $r_+$ is the smallest $r$ value for which $W_+=0$.
Close to the axis, we can expand $W_+$ treating $r$ as small 
\begin{equation}
\begin{split}
    W_+ = & \ \frac{(m-\ell)^2}{r^2}-2(1-g_{12}n_{2,0}-\omega) + g_{12}n''_{2,0} r^2 \\
    & + \left(4 n''''_{1,0} + 2n''''_{2,0}\right)r^4/4! + \mathcal{O}(r^6), \\
\end{split}
\end{equation}
where we have expanded $n_{1,2}$ up to and including quartic order terms.
Since the small $r$ behaviour of component 1 is $n_1\sim r^{2\ell}$, we assume at this stage we have an $\ell=2$ vortex so that $n''''_{0,1}\neq0$. To perform the calculation for $\ell\neq2$, one would need to retain larger terms in the $r$ expansion.
The unstable mode for the doubly quantised vortex occurs for $m=\ell=2$, hence we can write $W_+$ in the form,
\begin{equation}
\begin{split}
     W_+ = & \ 2\omega - 2U - \alpha r^2 + \beta r^4, \quad U = 1-g_{12}n_{2,0} \\
     \alpha = & \ -g_{12}n''_{2,0}, \qquad \beta = \left(4 n''''_{1,0} + 2n''''_{2,0}\right)r^4/4!
\end{split}
\end{equation}
where $\alpha$ and $\beta$ are positive constants.
For small enough $\eta$ the minimum of $W_+$ will be located in the region where this expansion is valid.
The minimum is $r_0=\sqrt{\alpha/2\beta}$ and the value of the potential there is $W_0=2(\omega-\omega_0)$ where $\omega_0=U+\alpha^2/8\beta$ is the maximum frequency which propagates in the cavity and is positive for small $\eta$.
Now we evaluate the cavity integral for $\omega$ close to $\omega_0$.
For this we expand $W_+$ this time around $r_0$,
\begin{equation}
    W_+ = W_0 + 2\alpha(r-r_0)^2.
\end{equation}
In this approximation $r_\pm=r_0\pm\sqrt{-W_0/2\alpha}$.
We can then compute $I$,
\begin{equation}
\begin{split}
    I = & \ \int^{r_+}_{r_-}dr\sqrt{-W_0-2\alpha(r-r_0)^2}, \\
    = & \frac{-W_0}{\sqrt{2\alpha}}\int^1_{-1}dz\sqrt{1-z^2} = -\frac{\pi W_0}{2\sqrt{2\alpha}},
\end{split}
\end{equation}
which can be written as,
\begin{equation}
    I(\omega) = \frac{\pi(\omega_0-\omega)}{|2g_{12}n''_{2,0}|^{1/2}} + \mathcal{O}[(\omega_0-\omega)^2],
\end{equation}
as stated in \eqref{Iaprx} of the main text.
It is also instructive to write $\omega_0$ as,
\begin{equation}
    \omega_0 = 1-g_{12}(n_{0,2}+n''_{0,2}r_0^2/4),
\end{equation}
where the term in parentheses is slightly smaller than the value of $n_2$ at $r_0$.
Since $n_2$ increases with $\eta$, we deduce that $\omega_0$ decreases as $\eta$ is increased from $0$.

\end{document}